\documentclass[12pt]{aastex62}

\begin{document}

\parindent=1.0cm

\title{Old Dogs, New Tricks: Re-examining Photographic Spectra of 
Plaskett's Star}

\author{T. J. Davidge}

\affiliation{Dominion Astrophysical Observatory,
\\Herzberg Astronomy \& Astrophysics Research Center,
\\National Research Council of Canada, 5071 West Saanich Road,
\\Victoria, BC Canada V9E 2E7\\tim.davidge@nrc.ca; tdavidge1450@gmail.com}

\begin{abstract}

	Photographic spectra of Plaskett's Star (PS; HR2420, HD47129, 
V640 Mon) that were recorded at the Dominion 
Astrophysical Observatory (DAO) have been 
digitized with a flatbed scanner. Many of 
the spectra were recorded during campaigns in 1922 and 1937, and
sample wavelengths between 0.39 and $0.50\mu$m. Spectra of poor quality 
are identified. Mean spectra near orbital phases 0.25 and 0.75 match many 
characteristics of synthetic spectra, although H$\gamma$ and \ion{He}{1} 
$\lambda 4388$ are exceptions. Evidence is presented that H$\gamma$ was 
affected by transient activity in 1937, but not in 1922. 
Emission lines of \ion{N}{3} and \ion{He}{2} 
move with wavelength in a manner that is consistent with them tracking 
the motion of the secondary, indicating that 
an 'f' spectral type designation should be assigned to the 
secondary. The location of the peak that is associated with the secondary 
in cross-correlation functions changes with time near phase 0.75, 
although the mean amplitude of the radial velocity curve of the secondary did 
not change between the two campaigns. There is also an offset in velocities of 
the primary measured from H$\gamma$ and \ion{He}{1} $\lambda 4472$ near 
phase 0.25. The velocity curves of the components 
suggest a mass ratio that is larger than 
previous estimates, although uncertainties associated with 
the spectroscopic features attributed to the secondary, coupled with the 
wavelength resolution of the spectra, complicate efforts to 
determine robust masses. We conclude that the peculiarities in the 
radial velocity curves of the components have thus been in place for over 
a thousand orbital cycles.

\end{abstract}

\section{INTRODUCTION}

	In a pioneering paper that is one of the first discussions 
of a non-solar stellar spectrum in the published literature, 
\citet{dra1877} examined absorption lines in a spectrum of Vega 
that was recorded on a photographic plate. While other highly innovative 
means of capturing astronomical spectra were implemented over the 
next century \citep[e.g.][]{oke1969, robandwam1972}, photographic plates 
remained the most commonly used detectors for spectroscopic observations 
until the widespread use of CCDs in the 1980s. As a result, 
photographic spectra are often the only means of 
examining the spectroscopic properties of stars prior to circa 1980. 

	There are a number of well-known shortcomings inherent to 
photographic plates, including low quantum efficiency, non-linearity, 
the inherent fragility of glass, and the vulnerability of the emulsion to 
damage. Still, the non-digital nature of photographic material is arguably the 
most fundamental impediment to their use. To date, the digitization of 
plates has largely been done with specialized machines that operate 
in carefully controlled environments. Limited access to these devices, 
combined with an inherently inefficient scanning process, can frustrate and 
complicate the digitization of photographic observations.

	The situation has changed with the availability of compact, 
high-quality flatbed scanners. While earlier generations of scanners 
failed to capture the full information content of 
photographic spectra \citep[e.g.][]{sim2009, dav2024}, this has changed with 
the current generation of top-of-the-line models. A pragmatic 
demonstration of this is that stellar spectra digitized with an 
Epson 12000-XL scanner compare favorably with those recorded 
with CCDs \citep[e.g. Figures 10 and 11 of][]{dav2024}. The use of 
flatbed scanners greatly simplifies the digitization process -- with 
a comparatively modest financial investment, it is now possible to capture 
efficiently the information content of photographic 
spectra in an office environment. 

	The Dominion Astrophysical Observatory (DAO) hosts a 
large collection of photographic spectra that were recorded 
on its telescopes over much of the past century. The largest number 
of plates were recorded with the original Cassegrain spectrograph 
on the 1.8 meter telescope \citep[]{pla1918}, starting in 
1918, while the Coud\'{e} spectrograph on the 1.2 meter telescope 
\citep[]{ric1968} has been in use since 1962. 
Many stars have been observed numerous times with these 
facilities, resulting in datasets that can be used to study 
long-term behaviour over timescales of many decades. 

	In the current paper we examine 
digitized spectra of the binary system Plaskett's Star 
(V640 Mon, HR2420, HD47129; hereafter PS). PS was targeted 
early in the operational life of the 1.8 meter telescope as part of a survey 
of emission-line O stars, and it was soon found to be a double 
line spectroscopic binary (SB2). A subsequent $\sim 1$ year observing campaign 
revealed velocity variations with a $\sim 14$ day period that 
were indicative of massive components \citep[]{pla1922}. At the time, PS was 
the most massive system known.

	While more massive systems have since 
been identified (e.g. Melnick 34 and HD38282, both in the LMC), 
PS remains one of the most massive binary systems to be detected 
in the Galaxy, and our knowledge of the basic properties of its components, 
including spectral types, luminosities, effective temperatures, surface 
gravities, and rotation velocities continues to evolve. 
Isolating spectroscopic signatures of the primary (defined here to be 
the brighter star at optical wavelengths) and the secondary (the fainter 
star) are key to determining their properties. Unfortunately, the 
spectrum of PS contains a mix of absorption and emission lines that 
appear to originate from a number of sources, and the complex nature of 
the composite spectrum confounded early efforts to isolate the spectrum 
of the secondary. 

	Emission associated with the Balmer and He lines was first noted by 
\citet{pla1922} and \citet{str1948}. The latter concluded that many of the 
weaker lines in the PS spectrum are non-stellar, and suggested that 
emission from a gas stream may contribute significantly to the light. 
\citet{stretal1958} noted that the weaker lines appear 
not to follow the orbital motion of the secondary, and 
discussed evidence for a shell, arguing that 
\ion{He}{1} lines that are blue-shifted by $\sim 700$ km/sec
originate from such a structure. \citet{hutandcow1976} found that 
the emission associated with \ion{He}{2}
$\lambda 4686$ may not be related kinematically to either star, while 
N emission at 4485 and 4503\AA\ may move with the primary. 

	Given that the components appear to be massive early-type stars 
then winds likely play a significant role in shaping the 
intrasystem and circumsystem environments. The ratio 
of the x-ray to bolometric flux from PS is similar to that 
found in O star binaries that likely contain colliding winds \citep[]
{linetal2006}. Some of the lines in the UV that are attributed to the secondary 
have P Cygni profiles, indicating an extended circumstellar region 
\citep[]{bagetal1992}. The properties of the intercomponent 
region may largely be defined by the wind from the secondary, 
which may dominate over that of the primary \citep[]{wigandgie1992}. 

	Variability in spectroscopic characteristics further 
complicates understanding the components of PS. \citet{stretal1958} found that 
the spectrum changes with time, with 'large and erratic flucuations' in the 
weaker lines that are attributed to the secondary. \citet{linetal2006} detect 
variations in the strengths of N and Fe lines in the X-ray regime, 
signaling that variability occurs over a range of wavelengths, 
species, and excitation conditions. There are also subtle line strength 
variations that may not be tied directly to 
the orbital properties of the system. The wings of 
H$\alpha$ emission change on timescales of $\sim 2$ days 
\citep[]{wigandgie1992}, suggesting that the intracomponent 
medium is structured. \citet{palandrau2014} identified two 
sub-orbital frequencies in spectroscopic properties, and conclude that 
these might originate from the secondary, the primary, or both. 
Non-radial pulsations, tidal interactions, and magnetic confinement of 
winds were considered as possible mechanisms for this variability, although 
Palate \& Rauw conclude that none of these explanations is problem-free.

	There are indications that the stars in PS have interacted, 
and if there has been large-scale mass transfer then this could 
complicate efforts to isolate the spectroscopic 
signatures of the components. For example, one expected consequence of mass 
transfer is a rapidly rotating secondary due to a gain in angular momentum 
from the primary, and evidence for such rapid rotation 
is seen \citep[e.g.][]{str1948, bagetal1992, bagandbar1996, gruetal2022}, 
including a flattening of the secondary in the orbital plane 
\citep[]{linetal2008}. Not only does rotation smear lines and make 
them harder to detect, it may also complicate efforts to assign 
a luminosity class due to gravity darkening \citep[e.g.][]{linetal2008}. 
A disk may also form that veils the host star. 

	There are chemical signatures indicating that the primary has evolved 
off of the ZAMS. Models of massive systems discussed by 
\cite{senetal2022} find that N is enhanced in a donor 
after the onset of Roche lobe overflow (RLOF), and that 
the He content increases as mass is lost. In contrast, N and He 
enhancement may be modest in the accretor due to rotational mixing. 
\citet{bagandbar1996} and \citet{linetal2006, linetal2008} find that the 
primary has non-solar abundance ratios, being deficient 
in C and O, while N is enhanced, which is consistent with CNO processing. 
\citet{linetal2008} also find that the secondary 
is overabundant in He and deficient in N. They note that the chemical abundances 
of both stars are consistent with post-RLOF evolution, with the rapid 
rotation of the secondary affecting its N abundance.

	If there have been interactions between the components 
then their properties should differ from those of 
isolated stars. The large-scale transfer of mass due to RLOF 
is expected to leave a primary that is substantially more luminous than would 
otherwise be expected for its present-day mass \citep[e.g.][]{senetal2022}, 
exacerbating the difference in luminosity that might have once existed 
between the two stars. \citet{bagetal1992}, \citet{bagandbar1996} and 
\citet{linetal2008} conclude that the ratio of fluxes in 
the blue and visual is $\sim 0.5$. Despite 
this difference, the stars appear to have similar 
spectral types \citep[e.g.][]{hutandcow1976}. \citet{bagetal1992}
decouple the spectra of the components to reveal 
spectral types O7.5I and O6I, while \citet{linetal2008} assign spectral types 
O8 III/I and O7.5III.

	Another complicating factor is that some of the motions detected 
in the spectra may not be associated with the orbital motions of the 
components, and this is compounded by the challenges isolating the spectrum 
of the secondary. Given the difficulties detecting the spectrum of the 
secondary, \citet{pla1922} used the radial velocities of the primary to obtain 
the orbital elements of the system, with the radial velocities of the secondary 
used only to obtain the amplitude of the velocity variations of that star. 
He found a mass ratio (i.e. the mass of the 
secondary divided by that of the primary) $\sim 0.8$ with a non-zero orbital 
eccentricity. 

	Subsequent studies confirmed that the radial velocities attributed to 
the components may not faithfully track their orbital 
motions. Based on 73 spectra that covered two orbital cycles, \citet{str1948} 
noted that the radial velocity curve of the primary changes between 
cycles, and that the orbital eccentricity obtained from the velocity curve 
varies with time. The radial velocity curve of the secondary was 
offset from that predicted by Plaskett's solution by $\sim$ 100 km/sec. 

	The problematic nature of the radial velocity curve of the secondary is 
not restricted to older datasets, where there might be difficulties 
isolating the centroids of weak, broad features. \citet{sti1997} examined 
cross-correlation functions (CCFs) based on UV spectra at 
different orbital phases, and concluded that the motion 
attributed to the secondary 'makes no sense', as the position of the peak in 
the CCF that is associated with the secondary does not 
define the same rest velocity as the primary. In addition, 
\citet{linetal2008} found that while the velocities measured for the 
primary over a wide range of wavelengths are consistent 
with each other, the agreement among velocities measured for the secondary 
is not as good. 

	\citet{gruetal2022} conclude that the rapidly rotating secondary is a 
magnetic star, and found that the Stokes V profiles are consistent with radial 
velocity variations that are substantially smaller than those found previously. 
More recent spectropolarimetric observations discussed by 
\citet{staetal2024} suggest that the radial velocity 
variation of the secondary may be even smaller than 
the 30 km/sec variation found by \cite{gruetal2022}.
The velocity of the secondary notwithstanding, 
it must have a mass of at least 12.6M$_\odot$. 
\citet{sti1987} combined velocities measured for 
the primary in the UV with all other published velocities at 
blue wavelengths to obtain a 'grand solution', and a mass function 
of 12.6 M$_{\odot}$ was found for a circular orbit. While the inclination 
of the system is uncertain, Stickland estimated that the secondary has a 
mass of at least 60 M$_{\odot}$. \citet{bagetal1992} subsequently 
combined additional velocity measurements with those considered by 
\citet{sti1987}, and found a mass ratio $1.18 \pm 0.12$. This is in contrast to 
most previous work, as it suggests that the secondary is more 
massive than the primary, as would be expected if 
there has been mass transfer. Based largely on orbital information defined by 
the primary, \citet{linetal2008} estimate 'best' masses of 55 and 57 
M$_{\odot}$ assuming an inclination of 71 degrees. 

	PS is a photometric variable, and the light curve provides 
additional clues into the nature of the system. When phased according 
to the orbital period, the light curve is asymmetric 
with an amplitude of a few hundredths of a magnitude \citep[]{mahetal2011, 
gruetal2022}. This modest amplitude is evidence that the structural properties 
of PS have changed considerably since the termination of mass transfer, as 
the lack of ellipsoidal variations of even moderate amplitude suggests 
that the components are not distorted by close proximity to 
their critical Roche surfaces. Still, interactions between 
the components due to stellar winds are not precluded, and the 
photometric variations may be due to activity on the 
primary that could be the product of shock-heated material. 
\citet{mahetal2011} found other periodicities in the light curve 
that are attributed to non-radial pulsations. The photometry produces 
tight light curves when phased to a period that is much shorter 
than that attributed to orbit motions, with a range in the 
periods deduced from different datasets \citep{gruetal2022}.

	The stellar content in the vicinity of PS contains 
additional clues into its age and evolution. 
Astrometric information obtained from the GAIA 
DR3 \citep[]{gai2016, gai2022} indicates that it is associated with 
Mon OB2 and the young cluster NGC 2244 \citep{dav2022}, 
which has an age $\sim 5.5$ Myr \citep[]{bonandbic2009}. 
However. this is not an ironclad age estimate as star formation in a cluster 
environment may proceed over many millions of years \citep[e.g.][]{biketal2019}.
Indeed, while \citet{dav2022} finds an entourage of stars that are 
associated with PS, none appear to be pre-main sequence objects 
\citep[]{linetal2008}, suggesting that PS is not in a star-forming 
environment at present. 

	In this paper we discuss photographic spectra of PS 
recorded with DAO telescopes that have been digitized with a flatbed scanner. 
The intent is to explore the information that can be 
extracted from the spectra, and demonstrate that the plates are 
more than curious historical artifacts from a bygone era. As the spectra 
sample a five decade timeframe, and thus cover more than one thousand orbital 
cycles, they provide a unique perspective on the spectroscopic properties 
of PS throughout much of the past century, prior to the advent of modern 
detectors. Many of the spectra were recorded during campaigns 
that were conducted over a single observing season, and in some cases 
this allows for a comparison of properties between consecutive orbital cycles.

\section{PLATE SELECTION AND CHARACTERISTICS}

	All plates of PS in the DAO collection were digitized, and a list of 
these is provided in Tables 1 (1.8 meter observations) 
and 2 (1.2 meter observations). The plate number, date of 
observation, orbital phase and other information are listed. The orbital 
phase was computed from the ephemeris in Table 2 of \citet{linetal2008}, 
where phase 0 is primary inferior conjunction. 
To the best of our knowledge the 1.8 meter spectra recorded in 
the 1930s and later have not been discussed in the published literature.

\startlongtable
\begin{deluxetable}{ccccc}
\tablecaption{Plate Information: 1.8 meter}
\tablehead{Plate \# & Date & HJD & Orbital & Notes\tablenotemark{a}\\
 & (YYYY MM DD) & -- 2400000 & Phase\tablenotemark{b} & \\}
\startdata
6858 & 1921 12 01 & 23025.967 & 0.361 & \\
6906 & 1921 12 16 & 23040.949 & 0.402 & \\
6953 & 1921 12 29 & 23053.838 & 0.297 & \\
6960 & 1922 01 01 & 23056.874 & 0.508 & \\
7006 & 1922 01 12 & 23067.849 & 0.270 & short camera\\
7095 & 1922 01 29 & 23084.802 & 0.448 & short camera\\
7138 & 1922 02 18 & 23104.753 & 0.834 & \\
7142 & 1922 02 19 & 23105.660 & 0.897 & \\
7161 & 1922 02 20 & 23106.693 & 0.969 & \\
7168 & 1922 02 20 & 23106.802 & 0.976 & short camera\\
7193 & 1922 02 23 & 23109.733 & 0.180 & \\
7209 & 1922 02 24 & 23110.632 & 0.242 & \\
7213 & 1922 02 24 & 23110.736 & 0.249 & \\
7217 & 1922 02 25 & 23111.675 & 0.315 & \\
7232 & 1922 02 26 & 23112.619 & 0.380 & \\
7234 & 1922 02 26 & 23112.785 & 0.392 & \\
7243 & 1922 02 27 & 23113.647 & 0.452 & \\
7264 & 1922 02 28 & 23114.769 & 0.530 & \\
7280 & 1922 03 07 & 23121.680 & 0.010 & \\
7282 & 1922 03 10 & 23124.689 & 0.219 & \\
7284 & 1922 03 12 & 23126.667 & 0.357 & \\
7289 & 1922 03 16 & 23130.640 & 0.632 & \\
7294 & 1922 03 16 & 23130.780 & 0.642 & short camera\\
7301 & 1922 03 18 & 23132.635 & 0.771 & \\
7302 & 1922 03 18 & 23132.808 & 0.783 & \\
7304 & 1922 03 19 & 23133.633 & 0.840 & \\
7330 & 1922 03 22 & 23136.697 & 0.053 & \\
7335 & 1922 03 23 & 23137.667 & 0.120 & \\
7385 & 1922 03 31 & 23145.661 & 0.675 & \\
7392 & 1922 04 03 & 23148.670 & 0.884 & \\
7403 & 1922 04 04 & 23149.664 & 0.953 & \\
7412 & 1922 04 05 & 23150.658 & 0.022 & \\
8334 & 1922 11 11 & 23370.027 & 0.260 & damaged\\
21768 & 1933 11 05 & 27382.047 & 0.945 & \\
21794 & 1933 12 03 & 27409.917 & 0.881 & alignment? \\
23677 & 1935 02 03 & 27836.799 & 0.498 & alignment? \\
26180 & 1936 10 08 & 28449.999 & 0.128 & \\
26211 & 1936 10 15 & 28457.019 & 0.616 & \\
26236 & 1936 10 17 & 28459.062 & 0.758 & \\
26282 & 1936 10 25 & 28467.033 & 0.311 & alignment? \\
26304 & 1936 11 01 & 28474.032 & 0.797 & bad \\
26329 & 1936 11 07 & 28480.082 & 0.218 & bad \\
26367 & 1936 11 21 & 28494.051 & 0.188 & \\
26376 & 1936 11 24 & 28497.026 & 0.395 & bad \\
26399 & 1936 11 26 & 28499.004 & 0.532 & bad \\
26400 & 1936 11 26 & 28499.067 & 0.536 & alignment? \\
26506 & 1937 01 02 & 28535.838 & 0.091 & \\
26517 & 1937 01 06 & 28539.837 & 0.368 & \\
26525 & 1937 01 07 & 28540.759 & 0.432 & \\
26529 & 1937 01 08 & 28541.765 & 0.502 & \\
26541 & 1937 01 11 & 28544.881 & 0.719 & alignment? \\
26568 & 1937 01 19 & 28552.854 & 0.272 & \\
26594 & 1937 01 31 & 28564.802 & 0.102 & \\
26626 & 1937 02 14 & 28578.796 & 0.075 & \\
26640 & 1937 02 25 & 28589.790 & 0.838 & alignment? \\
26646 & 1937 02 28 & 28592.650 & 0.037 & \\
26649 & 1937 03 03 & 28595.736 & 0.251 & \\
26654 & 1937 03 06 & 28598.674 & 0.455 & \\
26665 & 1937 03 15 & 28607.657 & 0.079 & \\
26666 & 1937 03 15 & 28607.690 & 0.082 & \\
26681 & 1937 03 16 & 28608.656 & 0.149 & \\
26694 & 1937 03 18 & 28610.730 & 0.293 & \\
26712 & 1937 03 29 & 28621.657 & 0.052 & alignment? \\
26718 & 1937 04 06 & 28629.687 & 0.610 & \\
26726 & 1937 04 07 & 28630.661 & 0.677 & \\
27391 & 1937 10 22 & 28829.056 & 0.458 & \\
27678 & 1938 03 07 & 28964.737 & 0.883 & red cut-off \\
28624 & 1938 10 06 & 29178.027 & 0.699 & \\
28992 & 1938 12 25 & 29257.906 & 0.247 & \\
29083 & 1939 02 23 & 29317.714 & 0.402 & \\
29090 & 1939 02 24 & 29318.835 & 0.480 & \\
A22101 & 1939 02 28 & 29322.704 & 0.748 & \\
A22112 & 1939 03 04 & 29326.639 & 0.022 & alignment? \\
A22113 & 1939 03 04 & 29326.681 & 0.025 & \\
A22114 & 1939 03 04 & 29326.737 & 0.028 & alignment? \\
A22124 & 1939 03 09 & 29331.684 & 0.372 & \\
A22125 & 1939 03 09 & 29331.738 & 0.376 & \\
29122 & 1940 03 23 & 29711.992 & 0.789 & \\
29174 & 1940 04 10 & 29729.673 & 0.017 & \\
29772 & 1940 11 18 & 29952.000 & 0.461 & \\
31966 & 1942 11 25 & 30688.948 & 0.651 & Wavelength calibration? \\
44460 & 1953 01 06 & 34383.740 & 0.300 & \\
47265+6 & 1954 11 07 & 35054.004 & 0.859 & Short camera, two spectra\\
\enddata
\tablenotetext{a}{Most spectra were recorded with the medium focus 
29\AA /mm camera, while a small number were recorded with the short focus 
55\AA /mm camera. The spectra with an 'alignment' entry 
in the last column have blurred lines and an uneven response with wavelength. 
It is suspected that these problems were due to unstable 
optical alignment in the spectrograph (see text, and Figure 5).} 
\tablenotetext{b}{From orbital elements in Table 2 of \citet{linetal2008}.}
\end{deluxetable}

	Some of the plates have a poor signal to noise ratio (S/N) and/or 
cosmetic flaws, such as scratches on the emulsion, fingerprints, and 
staining that occured during the development of the plate. Plates with 
cosmetic flaws are indicated in the fifth column, and these are most common 
among the 1.2 meter spectra recorded in 1974. Some of the spectra recorded for 
the 1937 campaign are affected by what appear to be optical alignment issues in 
the spectrograph, and these are discussed in more detail in 
Section 4. Whereas the centers of deep lines can be measured in those 
plates, their utility for the examination of line depth and strength is limited.

\begin{deluxetable}{ccccc}
\tablecaption{Plate Information: 1.2 meter}
\tablehead{Plate \# & Date & HJD & Orbital & Notes\tablenotemark{a}\\
 & (YYYY MM DD) & -- 2400000 & Phase\tablenotemark{b} & \\}
\startdata
4345 & 1969 03 07 & 40287.701 & 0.404 & \\
4419 & 1969 03 24 & 40304.671 & 0.583 & \\
5146 & 1969 10 23 & 40518.006 & 0.402 & \\
5386 & 1970 02 11 & 40628.757 & 0.095 & \\
6576 & 1971 02 07 & 40989.713 & 0.168 & \\
7352 & 1972 04 19 & 41426.699 & 0.522 & \\
7379 & 1972 05 05 & 41442.699 & 0.633 & \\
8804 & 1974 02 05 & 42083.885 & 0.172 & \\
8820 & 1974 02 20 & 42098.685 & 0.200 & \\
8825 & 1974 02 22 & 42100.844 & 0.350 & poor cosmetics \\
8907 & 1974 03 21 & 42127.684 & 0.214 & poor cosmetics \\
8908 & 1974 03 21 & 42127.718 & 0.216 & No arc\\
9329 & 1974 09 21 & 42312.012 & 0.018 & poor cosmetics \\
9421 & 1974 10 09 & 42329.927 & 0.262 & poor cosmetics \\
9467 & 1974 10 19 & 42339.979 & 0.961 & Long camera; weak lines \\
9480 & 1974 10 29 & 42349.928 & 0.652 & poor cosmetics \\
9481 & 1974 10 29 & 42349.960 & 0.654 & poor cosmetics \\
9545 & 1975 01 10 & 42422.938 & 0.723 & \\
9566 & 1975 01 29 & 42441.693 & 0.026 & \\
9584 & 1975 02 25 & 42468.658 & 0.899 & \\
9592 & 1975 02 26 & 42469.674 & 0.970 & poor cosmetics \\
9600 & 1975 03 06 & 42477.700 & 0.527 & low S/N \\
9607 & 1975 03 10 & 42481.679 & 0.803 & \\
9618 & 1975 03 11 & 42482.692 & 0.874 & \\
9641 & 1975 03 27 & 42498.681 & 0.985 & \\
\enddata
\tablenotetext{a}{Most of the spectra were recorded with the 32121 grating 
mosaic (10.1\AA\ /mm). Plates identified as having poor cosmetics are affected 
by non-uniformities in development and/or fingerprints.}
\tablenotetext{b}{From orbital elements in Table 2 of \citet{linetal2008}.}
\end{deluxetable}

	The 1.8 meter spectra were recorded through a slit with a 
prism as the dispersing element. The wavelength 
coverage extends from 0.39 to $0.50\mu$m, with the 
short wavelength cut-off largely defined by the throughput of the prism. 
This wavelength interval was targeted extensively 
in the early days of the DAO as it samples numerous features that 
are useful for velocity measurements \citep{pla1918}.
As these are slit spectra then the wavelength resolution and line profiles 
are susceptible to variations in the seeing. While seeing variations are to be 
expected at any ground-based observing site, they tend to be common in the 
DAO spectra of PS given that it has a meridian airmass of $\sim 1.4$ at DAO. 

	Most of the 1.8 meter spectra were recorded with 
the medium focus 28\AA/mm camera, although some were recorded with the 
short focus 55\AA/mm camera. The latter have a smaller physical size 
on the plate when compared with the former. However, the S/N and wavelength 
resolution of the digitized spectra from both cameras are similar, 
indicating that the characteristics of the camera optics are not limiting 
factors for defining the intrinsic wavelength resolution of these spectra.

	The 1.8 meter spectra from the 1920s and 1930s were recorded on 
Eastman 40 plates that were probably not hypersensitized. \citet{pla1921} 
discusses early negative experiences with hypersensitization at the DAO, 
leading us to suspect that the spectra recorded in 1922 -- 1923 were not 
hypersensitized. In section 4 it is shown that most of the plates recorded with 
the 1.8 meter at later dates have responses that are very
similar to the spectra recorded in the 1920s, suggesting 
that they also were not hypersensitized or, if our suspicions about 
hypersensitization are incorrect, they were hypersensitized with 
the same procedure used in 1922.

	The 1.2 meter spectra were recorded with an image slicer, 
which suppresses the potential for guiding errors while also improving 
throughput. The majority of these spectra were recorded with 
the 32121 grating mosaic (10.1\AA\ /mm), although some spectra were also 
recorded with the 3262 (9.0\AA\ /mm) and 3282 (6.5\AA AA\ /mm) 
gratings. The wavelength coverage of the 1.2 meter spectra typically 
extends from $0.35\mu$m to $0.55\mu$m, and we place emphasis on 
the 0.39 to $0.50\mu$m interval for compatability with the 1.8 meter spectra.

	The 1.2 meter spectra were recorded on Kodak type II or III plates 
with wavelength responses tuned to sample near-ultraviolet to visible 
wavelengths. \citet{mee1933, mee1935} discusses the various 
Kodak plate types used for spectroscopic applications.
The 1.2 meter plates were hypersensitized with either forming gas or Hg vapor. 

	Arc spectra bracket the PS spectrum on the 
vast majority of plates, and these are used for wavelength 
calibration. All of the 1.2 meter spectra as well as the 1.8 meter 
spectra recorded in the 1930s and at later dates have  
a calibration region to assess the photometric response. However, the signal 
tends to be saturated throughout much of this calibration region, and there 
is scattered light from saturated areas. In addition, 
unless the exposure time for the calibration signal 
is comparable to that of the stellar spectrum 
then reciprocity failure may also produce differences in response between 
the stellar spectra and calibration signal \citep[e.g.][]{mee1931}.
Given that the utility of this calibration information is unclear, 
then the intensity measurements of the spectra presented throughout 
this paper are based on the density of activated grains in the 
photographic emulsion. A differential approach for comparing spectra based 
on an assessment of response characteristics is thus adopted (Section 4). 

\section{SCANNING \& PROCESSING}

	The plates were digitized with an Epson 12000XL scanner. 
\citet{dav2024} demonstrated that the optical quality of this device is 
sufficient to recover the information content in spectra that were recorded 
with the 1.8 meter Cassegrain spectrograph. 
A scanning density of 2400 dpi was used, which corresponds to a 
spatial resolution of $\sim 10\mu$m. This is comparable to, or smaller 
than, the grain size of most photographic emulsions. 

	The digitized spectra were processed according to the procedures 
described by \citet{dav2024}. All processing was done with tasks in 
IRAF \citep[][]{tod1986, tod1993}. The processing included background 
subtraction using measurements near the plate edges to 
suppress signal from scattered light and plate fogging.

	 Unlike the spectra of Vega that 
were examined by \citet{dav2024}, the spectral resolution of the PS data 
typically does not change along the slit, and this is likely due to 
the longer exposure times for the PS spectra. Still, there are a 
modest number of exceptions that are likely due to seeing, and 
the manner in which the slit is illuminated by the wings of the 
point spread function. Experimentation found that signal that is 
within 20\% of the peak value of the mean illumination profile constructed 
by averaging the light distribution along the slit 
tended to have a stable spectral resolution. Therefore, 
signal within 20\% of the mean peak signal was extracted from each plate 
for further processing. In addition to stability in spectral 
resolution, this extraction criterion also restricts the final spectra to 
those parts of the light profile that have the highest S/N, while also reducing 
the effect of scattered light from the arcs that closely bracket the 
1.8 meter spectra. In contrast, there is a distinct gap between the 
arc and stellar spectra on the 1.2 meter plates. Residual 
signatures of bright arc lines are not evident in any of the extracted spectra.

	The signal in individual columns 
of each digitized spectrum was summed, and a wavelength 
calibration that was based on a linear interpolation of the arcs 
that bracket each spectrum was applied. The wavelength 
calibration was then placed in the heliocentric frame.
The wavelength calibrated spectra were normalized to a pseudo-continuum 
that was based on a polynomial fit to each spectrum. Recursive 
clipping was used to suppress absorption and emission lines when fitting 
the continuum. 

	Some 1.8 meter spectra recorded in the 
1930s appear to be affected by optical alignment 
problems that we suspect are due to flexure 
and mechanical issues in the spectrograph. Evidence for 
mis-alignment comes in the form of satellites and sub-structures 
in arc lines, as well as periodic undulations in the 
continuum that have an amplitude of up to $\sim 10\%$ and are likely 
due to scattered light. An effort was made to remove the variations in the 
continuum by fitting high-order polynomials, although 
this had mixed success and affected the wings of broad lines, such 
as H$\gamma$. These spectra are discussed further in Section 4.2.

\section{CHARACTERISTICS OF THE SPECTRA}

\subsection{General Properties}

	There are two sets of 1.8 meter spectra and one set of 1.2 meter 
spectra that were recorded over the course of a single observing season. 
For the remainder of the paper we refer to spectra recorded 
between {December 1 1921 and March 5 1922 as belonging to 
the 1922 campaign, while those obtained between October 8 1936 and March 7 
1937 belong to the 1937 campaign. Spectra recorded with the 1.2 meter between 
September 21 1974 and March 27 1975 belong to the 1975 campaign. The two 
1.8 meter campaigns provide insights into the properties of PS over a 15 year 
baseline. The 1.2 meter spectra extend the time coverage by another four 
decades, although many of those spectra are of poor quality (Table 2). 

	The average properties of spectra in each campaign 
have been examined to assess similarities and differences. 
Spectra that were recorded for each campaign 
were aligned on the centers of H$\gamma$ and \ion{He}{1} $\lambda 4471$, and 
the results were combined by taking the median signal at each wavelength. 
Spectra identified in Table 1 as having a poor S/N were not included. 
Given that (1) the primary star is the dominant contributor 
to the system light near 4400\AA\ , and (2) the alignment was 
done using features that are prominent in its spectrum, then the combined 
spectra are expected to be largely representative of the primary. Still, the 
high rotational velocity of the secondary means that there will likely 
be residual signatures from that star in each combined spectrum. 

\subsubsection{1.8 Meter Spectra}

	The median spectra constructed from the 1922 and 1937 campaigns are 
compared in Figure 1. The difference between these, in the sense 
1922 -- 1937, is shown at the bottom of the figure. At first glance, a 
comparison of the campaign spectra reveals 
few surprises. As expected for O stars, the deepest absorption features 
are those of H and He. In addition to absorption lines, there are 
also a number of emission features. Emission is present in the 
wings of H$\beta$, but is not obvious in the higher order 
Balmer lines. \ion{N}{3} emission at 4630 and 4634\AA\ and \ion{He}{2} 
emission at 4686\AA\ are also present. The N emission is a signature of the 'f' 
sub-class for O stars, while \ion{He}{2} $\lambda 4686$ in emission is 
indicative of a supergiant luminosity class \citep[e.g.][]{mar2018}. 

\begin{figure}
\figurenum{1}
\epsscale{1.0}
\plotone{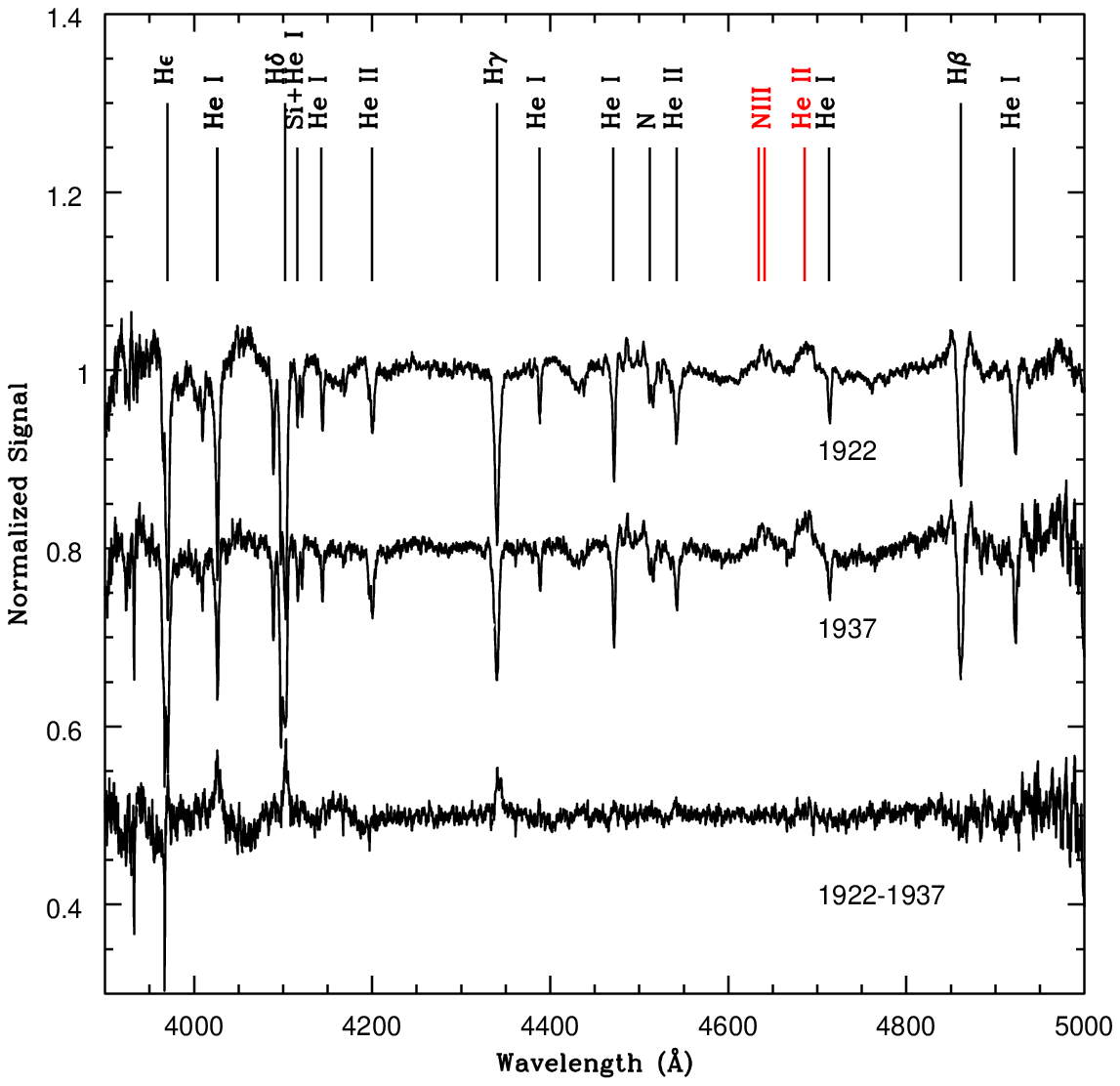}
\caption{Median 1.8 meter spectra of PS, constructed by taking the 
pixel-by-pixel median of individual campaign spectra after aligning to 
the rest wavelengths of H$\gamma$ and \ion{He}{1} 
$\lambda 4471$. As with all other figures 
in this paper, the vertical axis is a measure of the density of activated 
grains in the emulsion, and has not been corrected for departures from 
linearity. Given the relative brightnesses of the component 
stars, the median spectra are largely representative of the 
primary, as features associated with the secondary are suppressed or blurred. 
Still, some features that are associated with the secondary remain (see 
text). The S/N diminishes near the blue and red edges of the spectra due to the 
lower throughput at these wavelengths. The difference between the two spectra 
is shown at the bottom. The depths of the Balmer lines and \ion{He}{1} 
$\lambda 4026$ differ at the $\sim 5\%$ level. However, between 
4400 and 4800\AA\ , where the throughput is highest, 
the variance in the difference is $\sim \pm 1\%$. 
It is argued in the text that the difference in the depth of H$\gamma$ 
is not due to plate characteristics. (Note: The data used to construct this 
figure will be made available online in the published version of this paper).}
\end{figure}

	It is shown below that \ion{N}{3} emission moves with orbital phase 
in the same sense as the secondary. The \ion{N}{3} lines 
in Figure 1 are thus blurry residuals, and so do not 
represent these features as they would appear in individual spectra. 
The central absorption feature in the \ion{He}{2} emission profile 
also moves in a manner that is consistent with it belonging to the 
secondary, and this is contrary to what was found by \citet{hutandcow1976}. The broad DIB band near 4430 \AA , which is one of the 
deepest interstellar features in the blue - red wavelength region 
\citep[e.g.][]{her1975}, is also apparent.

	The overall throughput of both sets of 1.8 meter spectra is highest 
between 0.42 and $0.48\mu$m and, with the exception of H$\gamma$ (see below), 
there is good agreement between the median campaign 
spectra at these wavelengths, with a dispersion in the 
differenced spectrum between 0.44 and $0.48\mu$m of $\sim 
\pm 1\%$. The agreement between the median spectra at these wavelengths 
suggests that the plates recorded for the two campaigns tend to have similar 
response characteristics, at least among features that 
(1) have strengths that are within $\sim 10\%$ of the continuum, and (2) 
are the result of combining many plates to suppress environmental variations, 
such as differences in wavelength resolution due to seeing variations. 
The comparison in Figure 1 thus opens the possibility of making differential 
comparisons between the two sets of 1.8 meter spectra in the 0.42 -- $0.48\mu$m 
wavelength region, at least when considering differences on the order of 1\% 
or higher. This has broad implications for plates in the DAO collection 
as it suggests that spectra of other stars recorded at the DAO between 
1921 and 1938 should thus have similar response characteristics.

	H$\gamma$ appears to be a special case, as it is the one feature 
between 0.44 and $0.48\mu$m that changes with time. While the depths of 
H$\gamma$ in the spectra differ at the $\sim 5\%$ level, a similar difference 
is not seen for \ion{He}{1} $\lambda 4471$. We suspect that the 
difference in the depths of H$\gamma$ between the 
two spectra in Figure 1 is not an artifact of the response of the photographic 
plates, as H$\gamma$ and \ion{He}{1} $\lambda 4471$ are close together in 
wavelength, and the He line has an intrinsic depth that is not greatly 
different from that of H$\gamma$. These features should then sample similar 
parts of the characteristic curve of each plate. While these are slit spectra, 
the difference in H$\gamma$ depth is not due to guiding errors and seeing 
variations, as these would affect the characteristics of all lines in similar 
proportion, which is not the case in Figure 1. The behaviour of H$\gamma$ 
is examined further in Section 4.2. 

	The differences between the campaign spectra grow in size 
outside of the wavelength interval where the throughput is highest. 
H$\delta$ and \ion{He}{1} $\lambda 4026$ are stronger in the 1922 spectrum, 
differing from the depth in the 1937 campaign spectrum by between 5 and 10\%. 
As for H$\beta$, it is weaker in the 1922 spectrum, and in Section 4.2 
evidence is presented for variations in the shape of this line, likely due 
to emission. 

\subsubsection{1.2 meter Spectra}

	A subset of the 1975 campaign spectra were also median-combined, 
and the result is shown in Figure 2. Many of the PS plates recorded 
with the 1.2 meter have cosmetic flaws that are 
largely due to problems with the development process and plate 
handling (e.g. fingerprints). The median spectrum shown in 
Figure 2 was constructed from the five spectra that do not have 
cosmetic issues, which are plates 9545, 9566, 9584, 9607, and 9641. The 
spectrum that results if more spectra are combined has slightly 
different absorption line depths. 

\begin{figure}
\figurenum{2}
\epsscale{1.0}
\plotone{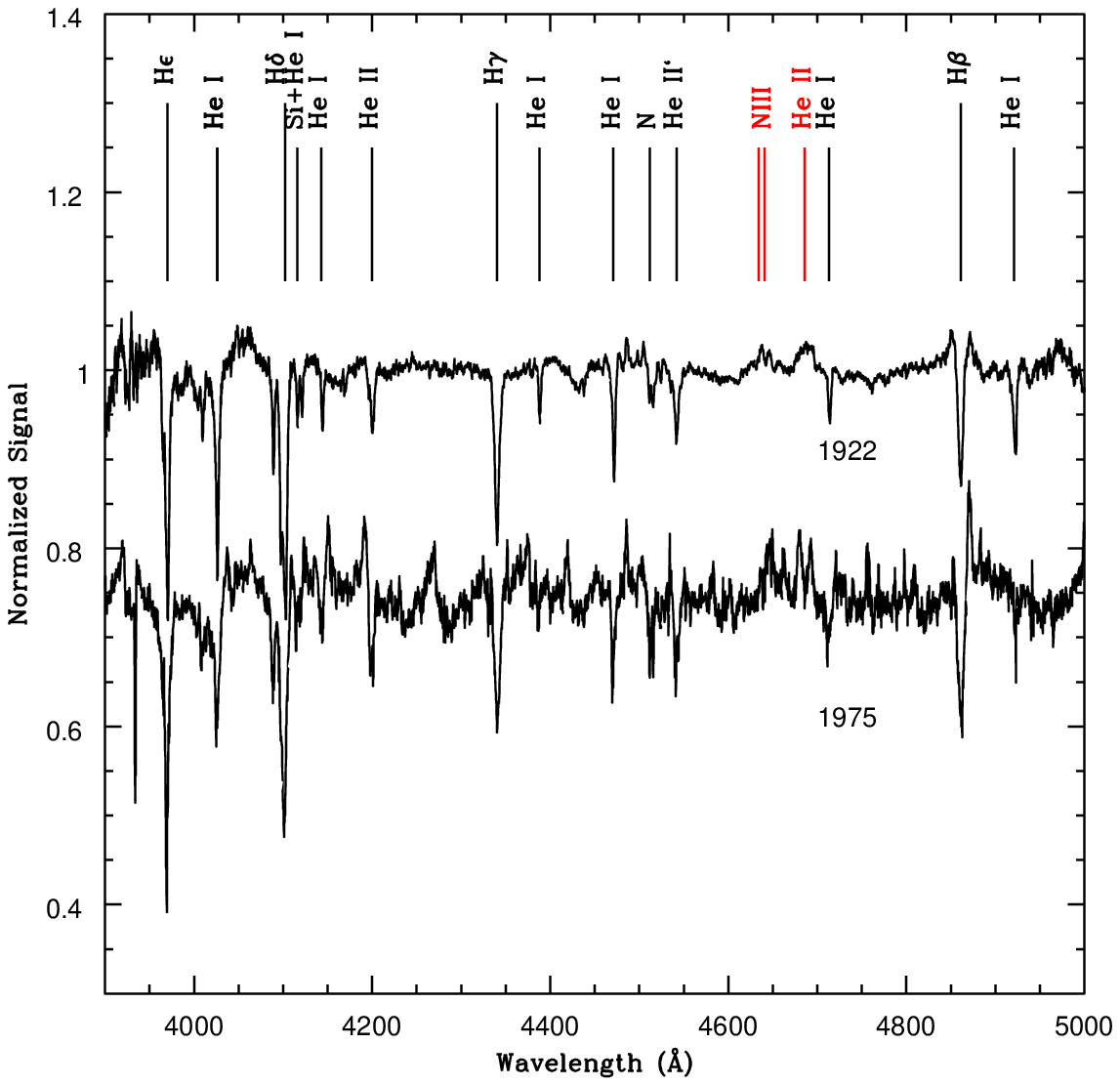}
\caption{Composite 1975 campaign spectrum of PS, constructed from the five 
spectra recorded during that campaign that do not have major cosmetic defects. 
As with the spectra in Figure 1, individual spectra were shifted 
to align H$\gamma$ and \ion{He}{1} $\lambda 4471$ with their rest 
wavelengths before taking the median at each pixel. 
The median 1922 campaign spectrum from Figure 1 is also shown. The 
differences between the 1975 and 1922 campaign spectra are similar to those 
between the 1937 and 1922 spectra in Figure 1. Still, while 
there is not a large difference in the widths of absorption lines, there is 
a noticeable difference in the properties of \ion{He}{2} $\lambda 4686$ 
emission, which appears as two distinct components in the 1.2 meter spectrum.
(Note: The data used to construct this figure will be made available 
online in the published version of this paper).}
\end{figure}

	A comparison of the 1975 and 1922 campaign spectra in Figure 2 
indicates that the spectral resolution of the 1.2 meter spectra, 
as gauged from the widths of absorption lines, 
is not greatly different from that of the 1.8 meter 
spectra. The largest difference in resolution is apparent in the 
character of \ion{He}{2} $\lambda 4686$ emission, which appears as two 
well-separated peaks in the 1.2 meter data. The general similarity of the 
1.2 meter and 1.8 meter campaign spectra is noteworthy given that the 1.2 meter 
plates were recorded more than three decades after the 1937 campaign, 
and so use different emulsions and hypersensitization techniques. 
In addition, the 1.2 meter spectra were recorded with an image slicer, 
and so the spectral resolution should be less susceptible to seeing 
variations than the 1.8 meter slit spectra.

\subsection{Consistency}

	While Figures 1 and 2 demonstrate that there is an overall consistency 
between DAO spectra recorded in the 1920s and 1930s when their average 
characteristics are considered, there are significant plate-to-plate variations 
in the characteristics of absorption features, and these are examined in this 
section. The intrinsic depths of features in the spectrum of 
an interacting binary change with orbital phase, due 
to variations in the observed brightness of each star as well as the 
motions of lines with wavelength during an orbital cycle. To suppress these, 
we compare spectra in narrow orbital phase intervals. 
The 1.2 meter spectra are not included in these comparisons given the 
limited number of cosmetically 'good' spectra. 

	The depths of prominent lines in the 1922 and 1937 campaigns 
in the orbital phase intervals 0.95 - 0.05 and 0.2 - 0.3 are compared in 
Figures 3 (H$\gamma$ and \ion{He}{1} $\lambda 4471$) and 4 (H$\beta$). 
H$\gamma$ and \ion{He}{1} $\lambda 4471$ are separated by only 130\AA\ , and 
are at wavelengths where the overall system throughput is highest. 
Barring intrinsic variations in line strengths in one or both of the 
components then H$\gamma$ and \ion{He}{1} $\lambda 4471$ should vary in 
lockstep.

\begin{figure}
\figurenum{3}
\epsscale{1.0}
\plotone{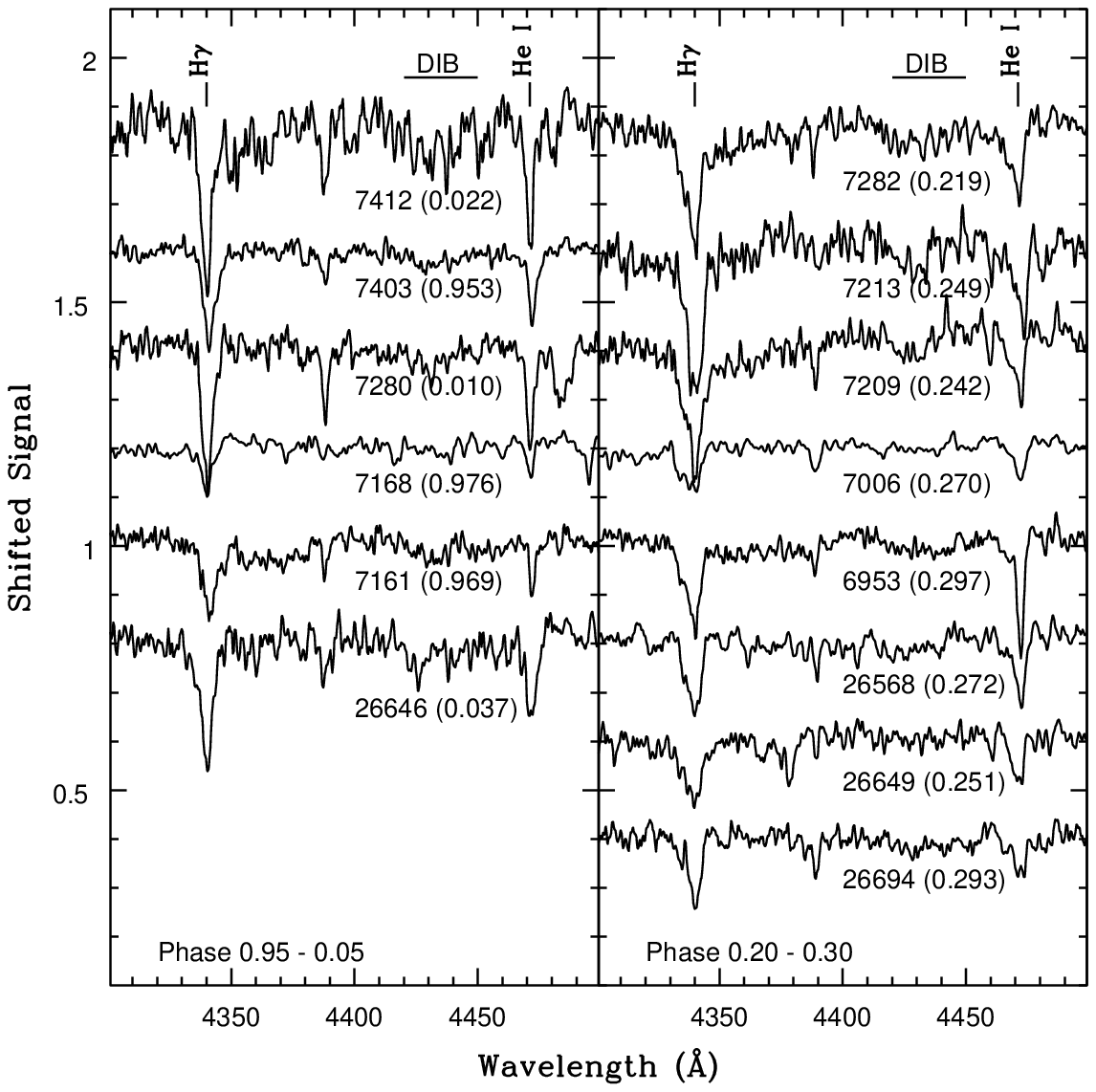}
\caption{Selected spectra in the wavelength interval 4300 to 
4500\AA\ that were recorded during the 1920s and 1930s. Orbital 
phases are in brackets next to the plate numbers. Spectrum-to-spectrum 
variations in the absolute and relative depths of H$\gamma$ and \ion{He}{1} 
$\lambda 4471$ are seen. The depths of H$\gamma$ and \ion{He}{1} $\lambda 
4471$ vary in sync in most cases, as expected if the range in line 
characteristics is not a result of intrinsic spectroscopic variations, but 
is instead due to observational effects, such as seeing variations. Still, the 
depths of H$\gamma$ and \ion{He}{1} $\lambda 4471$ in spectra 6953 and 26568 
are comparable, hinting at an intrinsic origin for the variation in 
some features. With the exception of spectrum 26646, the sharpest 
and deepest features are seen in the 1922 campaign spectra, 
and this is consistent with the relative depths of features in the 
median spectra in Figure 1.}
\end{figure}

\begin{figure}
\figurenum{4}
\epsscale{1.0}
\plotone{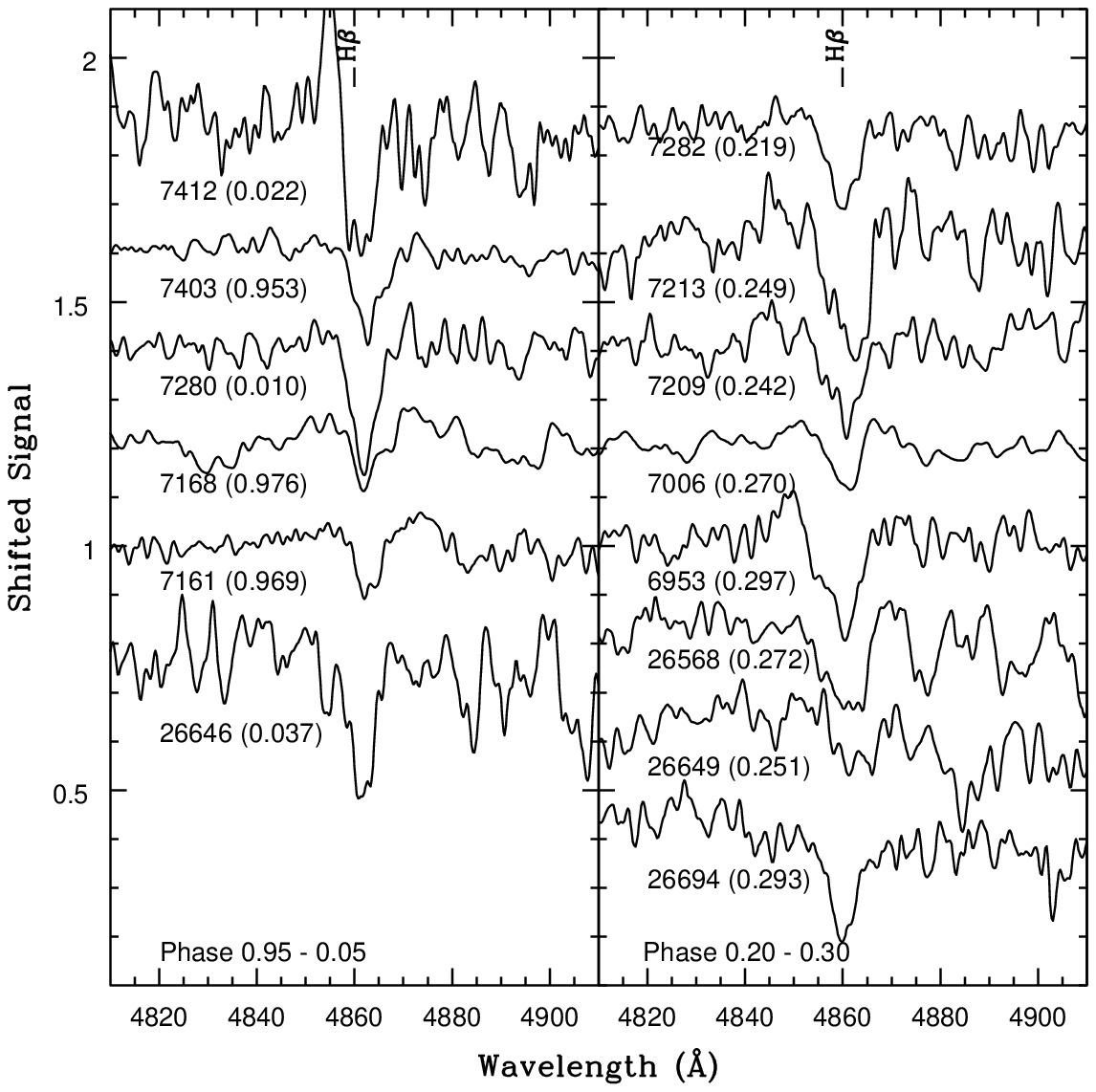}
\caption{Same as Figure 4, but showing H$\beta$. H$\beta$ 
is on the shoulder of the response curve where the overall throughput is 
markedly lower than near H$\gamma$ and \ion{He}{1} $\lambda 4471$, 
and so the spectra are noisier than in Figure 3. There is 
a clear dispersion in the depths and shape of H$\beta$, in the sense 
that the character of the line core varies with time, 
changing from a pointed to a flat or double-pointed morphology.}
\end{figure}

	The lines in Figures 3 and 4 have similar behaviour, in the 
sense that the spectra with the deepest lines in Figure 3
also have the deepest H$\beta$ line in Figure 4. There is an expectation that 
the relative depths of H$\gamma$ and \ion{He}{1} $\lambda 4471$ should vary in 
sync. However, while \ion{He}{1} $\lambda 4471$ is slightly shallower 
than H$\gamma$ in most cases, two exceptions are 
the spectra extracted from plates 6953 and 26568, where H$\gamma$ and 
\ion{He}{1} $\lambda 4471$ have similar depths. This is another indication 
that the depth of H$\gamma$ may be subject to variations that are 
intrinsic to the spectrum of the primary, rather than observing conditions.

	The comparisons in Figure 4 are compromised 
somewhat by the S/N, although there is 
a clear dispersion in the depth of H$\beta$. 
The shape of H$\beta$ changes with time, 
in the sense that there is a pointed line core at some 
epochs, and a flat core at others. Variations of this nature are seen 
in the spectra of evolved O stars. For example, the O supergiant 
HD192639 has a spectral type that is similar to 
that of the components in PS. H$\beta$ in the spectrum of this star has a 
variable morphology like that seen in Figure 4 \citep[]{rauandvre1998}.

	If large scale spectroscopic variations over the timescales 
of a few hours are absent in PS then spectra that are 
recorded on the same night are an important means of examining the consistency 
of the photographic spectra, and may provide insights into the nature of any 
plate-to-plate differences. Three spectra of PS were recorded with the 
1.8 meter on the night of March 4 1939, and these are plates A22112, A22113, 
and A22114. The spectra extracted from these plates are compared in Figure 5. 

\begin{figure}
\figurenum{5}
\epsscale{1.0}
\plotone{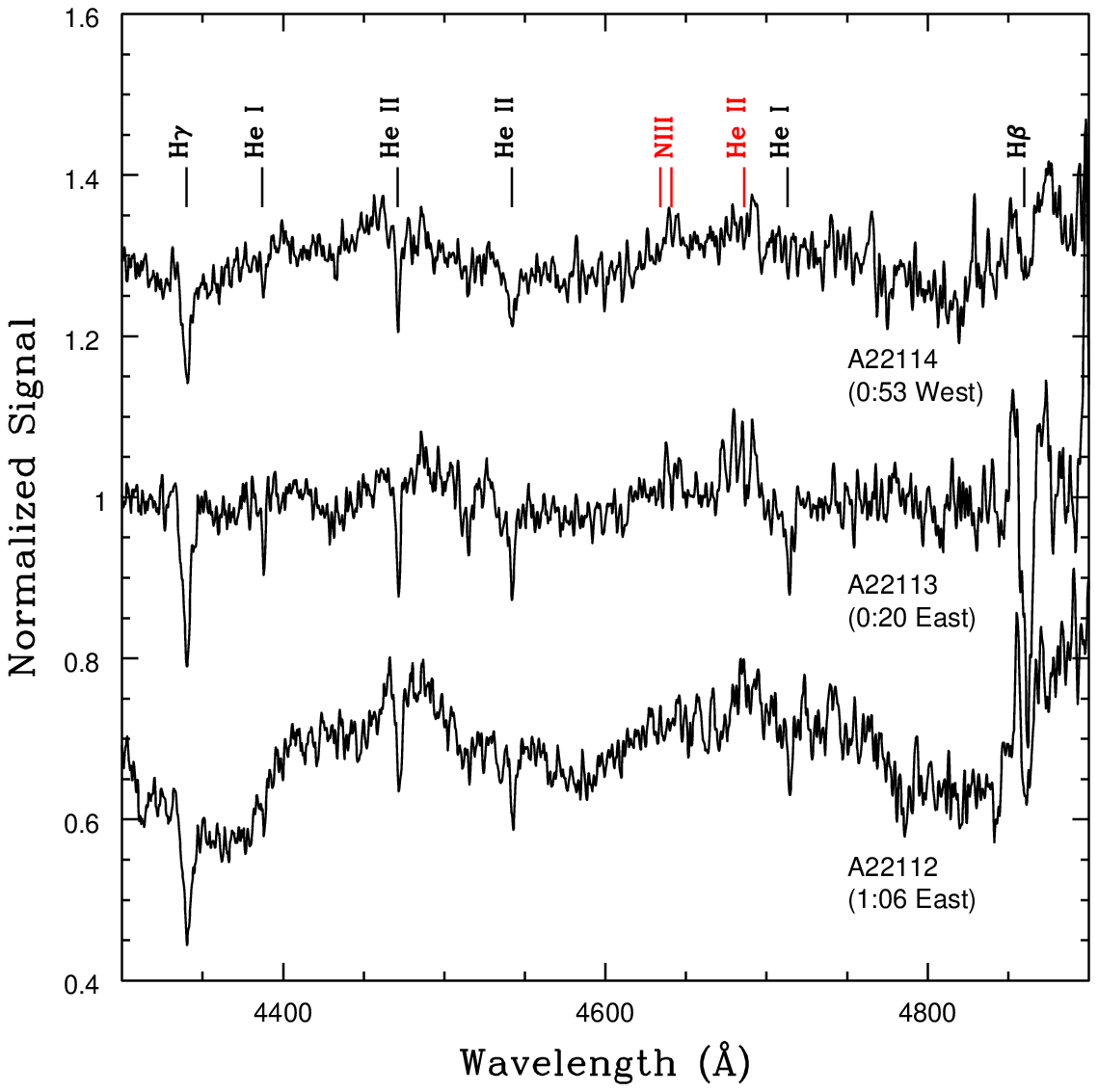}
\caption{Spectra recorded with the 1.8 meter telescope on the night of March 
4, 1939. The hour angle at the start of the exposures is shown in 
brackets; spectrum A22113 was recorded as PS transited the meridian. 
The spectra have been normalized to a high-order polynomial that 
was fit to the pseudo-continuum in an attempt to remove large-scale 
variations in the system response, while not affecting the 
depths of absorption lines. However, the highest order polynomial that could 
safely be applied to the A22112 and A22114 spectra left residual variations 
in the pseudo-continuum. These variations are periodic 
and synced in phase, suggesting a common 
origin. There is a tendency for the absorption lines to be broader and 
shallower in A22112 than in the other spectra, while the A22113 spectrum 
has a comparatively flat pseudo-continuum coupled with the sharpest 
and deepest lines. In contrast to plate A22113, the arc lines 
on the A22112 and A22114 plates have conspicuous satellites, 
suggesting that the variations in the pseudo-continuum are due to optical 
alignment issues, possibly arising from flexure in the spectrograph 
due to the orientation of the telescope ('flopture').} 
\end{figure}

	The hour angle at the start of each exposure is shown in brackets 
under the plate number in Figure 5. The A22112 spectrum was recorded 
one hour to the east of the meridian and has an uneven 
continuum. Similar, but lower amplitude variations, are seen 
in the A22114 spectrum, which was recorded one hour to the west of the 
meridian. The uneven nature of the continuum is a feature that is shared with 
other spectra recorded during the 1930s, and these are indicated with the 
'alignment?' entry in the right hand column of Table 2. The A22113 
spectrum has the sharpest lines and the flattest continuum in Figure 5, 
and this spectrum was recorded while PS was crossing the meridian. 
While absorption lines are visible in all three spectra, 
signalling that they might be useful for velocity 
measurements, there are differences in the depths 
and widths of lines that appear to be correlated with the amplitude of the 
continuum flucuations, and these frustrate studies of line profiles. 

	The spectrum-to-spectrum differences in Figures 3, 
4, and 5 are likely due to a number of factors. 
The spectra with poor S/N are easiest 
to explain. The sky transparency tends to be poor during the winter in 
Victoria, with variations in cloud cover occuring over 
short timescales. Exposures that were started in what were initially 
acceptable conditions might then be truncated prematurely, with only 
a fraction of the intended signal recorded.

	The 1.8 meter spectra were recorded through a slit, and so differences 
in line depth due to seeing and/or guiding errors might be expected. 
PS culminates on the sky at DAO near an airmass of 1.4. Given the 
roughly one hour exposure time required to obtain a photographic 
spectrum of PS, coupled with the relatively high airmass 
and awkward positioning of the telescope while observing PS and the need for 
manual guiding at the time these data were recorded, then 
the spectra are likely susceptible to seeing variations 
and -- possibly -- guiding errors. Both of these will affect the 
wavelength resolution of slit spectra, such as are apparent 
in Figures 3 and 4. Additional evidence that observational factors 
may dominate the dispersion in line properties was 
presented earlier in the paper, where the median campaign spectra, 
in which variations due to observing conditions are suppressed, were 
found to agree for most (but not all) lines. 

	The continuum variations in the A22112 and A22114 spectra 
are periodic and synchronized in wavelength. 
This is indicative of a common origin, and suggests that 
they are not due to plate flaws, issues 
with the development process, or seeing flucuations. 
Rather, we suspect that the continuum variations have an origin that is related 
to the spectrograph. The arc emission lines on these plates have satellites 
that are suggestive of an optical alignment issue, and the 
number of plates in Table 1 that are flagged with alignment issues in 
Table 1 suggest that this was not a rare occurence in the 1930s. Problems with 
optical alignment may occur if there is flexure in the spectrograph 
as the telescope tracks an object at high airmass. 
Flexure would be less of a problem when the target is close to the meridian. 
This is consistent with the behaviour of the spectra in Figure 5.

\subsection{Variations with Orbital Phase}

	Given the spectrum-to-spectrum differences discussed 
in the previous section, whenever possible we consider 
the means of spectra to examine changes in line properties with 
orbital phase. The averaging of spectra suppresses environmental 
factors, and so should provide a more homogeneous dataset when 
making differential comparisons. Systematic factors, such as the 
non-linear response of the photographic plate, are not suppressed.

	Mean spectra in the orbital phase ranges 0.15 -- 0.35 
and 0.65 -- 0.85 are shown in Figures 6 (1922 campaign), 7 
(1937 campaign), and 8 (1975 campaign). These phases 
sample the points in the orbit where the wavelength 
separation between the spectra of the components is greatest, 
and so should then provide the most reliable insights into their spectral types 
and relative luminosities. Three wavelength regions centered on H$\gamma$, 
H$\beta$ and the \ion{N}{3} emission lines are shown in Figures 6 and 7. 
The wavelength interval that covers H$\beta$ is not shown in Figure 8 
as the S/N of the 1.2 meter spectra at these wavelengths is poor. The 
motion of lines due to the orbital motions of the components are evident 
in all three figures, as are changes in line shape. The latter are due 
in large part to the orbital motion of the secondary, the spectrum of which 
moves with respect to that of the primary and forms a notch in the line 
profiles. 

\begin{figure}
\figurenum{6}
\epsscale{1.0}
\plotone{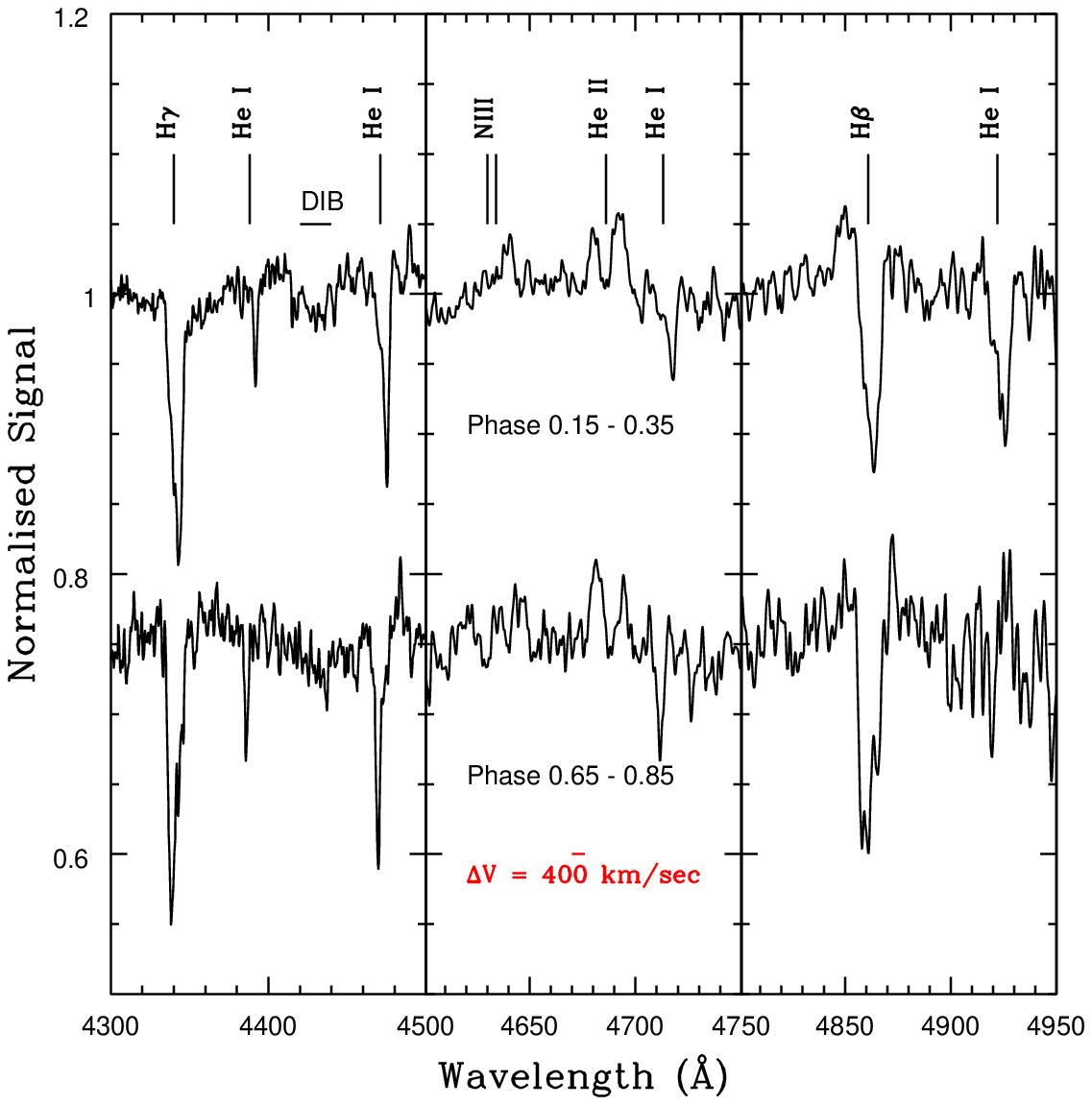}
\caption{Mean 1922 campaign spectra in the orbital phase 
intervals 0.15 - 0.35 and 0.65 - 0.85. The horizontal red line in the middle 
panel shows the maximum shift in wavelength due to the motion of 
the primary (i.e. $\sim 6$\AA). Shifts in the mean wavelengths of H and He 
absorption lines as well as changes in their profiles due to the motion of the 
secondary are seen. While \ion{N}{3} emission is weak 
between phases 0.15 and 0.35, it is more pronounced near phase 0.75. The 
character of \ion{He}{2} $\lambda 4686$ and the emission in the shoulders of 
H$\beta$ both change with orbital phase. The central 
absorption component of \ion{He}{2} $\lambda 4686$ and the H$\beta$ emission 
follows the motion of the secondary.}
\end{figure}

\begin{figure}
\figurenum{7}
\epsscale{1.0}
\plotone{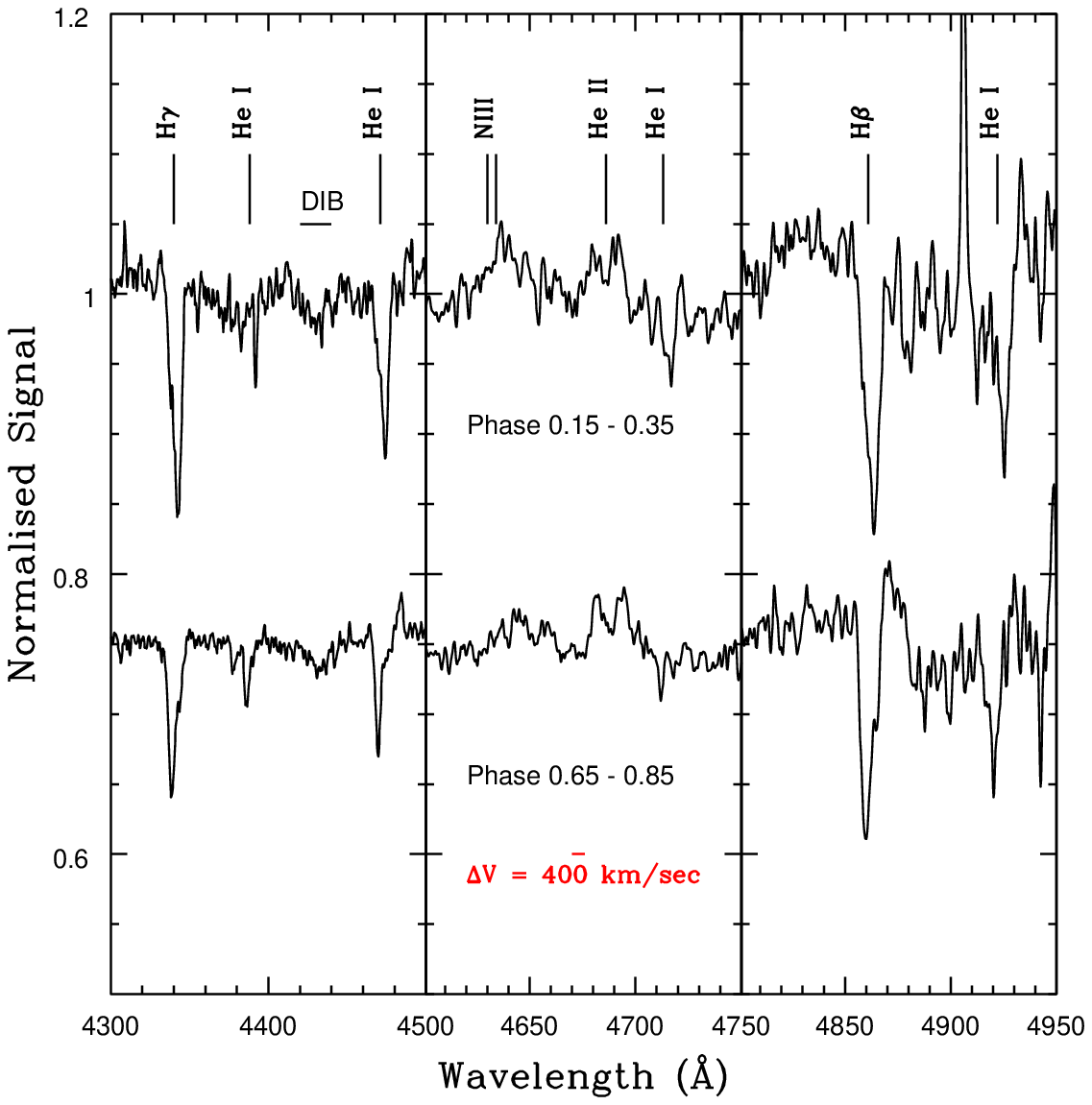}
\caption{Same as Figure 6, but comparing the 1937 campaign spectra. 
As in Figure 6 (1) the spectrum of the secondary is clearly evident in many of 
the absorption lines, (2) the character of \ion{N}{3} emission 
changes with orbital phase, and (3) the emission associated with 
\ion{He}{2} $\lambda 4686$ and in the shoulders of H$\beta$ tracks the 
motion of the secondary.} 
\end{figure}

\begin{figure}
\figurenum{8}
\epsscale{1.0}
\plotone{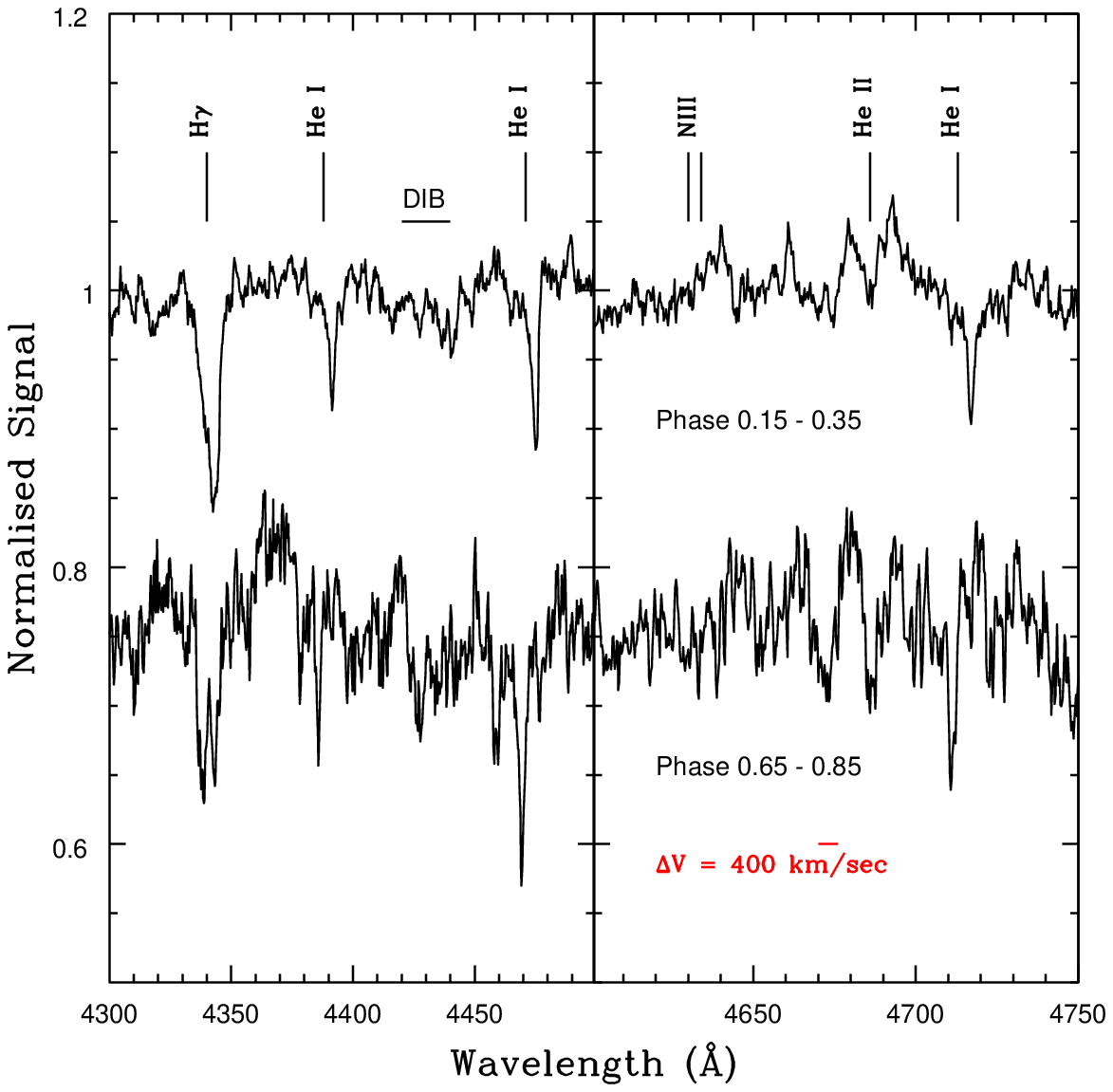}
\caption{Same as Figure 6, but comparing the 1975 campaign 
spectra. The wavelength interval covering H$\beta$ is not shown as 
the S/N is too low to permit a meaningful comparison between orbital 
phases. The behaviours of the absorption and emission lines are similar 
to that in Figures 6 and 7.}
\end{figure}

	The relative depths of H$\gamma$ and \ion{He}{1} $\lambda 4471$ is a 
measure of spectral type. The locations of the notch that is due to 
the secondary in the line profiles indicates 
that \ion{He}{1}/H$\gamma$ is smaller in the secondary 
than in the primary, suggesting that the secondary has a later spectral 
type than the primary. Moreover, the H$\gamma$ and \ion{He}{1} profiles 
near phases 0.25 and 0.75 do not mirror each other, in 
that the notch in the line profile that is attributed to the secondary 
near phase 0.25 is not seen at the same depth as that near phase 0.75. 
This is demonstrated in Figure 9, where the profiles of H$\gamma$ and 
\ion{He}{1} $\lambda 4471$ from Figures 6, 7, and 8 are overplotted. 
The mean spectrum from the 1937 campaign spectrum for phases 0.65 -- 
0.85 has been scaled by a factor of 1.5 in this figure to match the depth 
of H$\gamma$ between phases 0.15 and 0.35.

\begin{figure}
\figurenum{9}
\epsscale{1.0}
\plotone{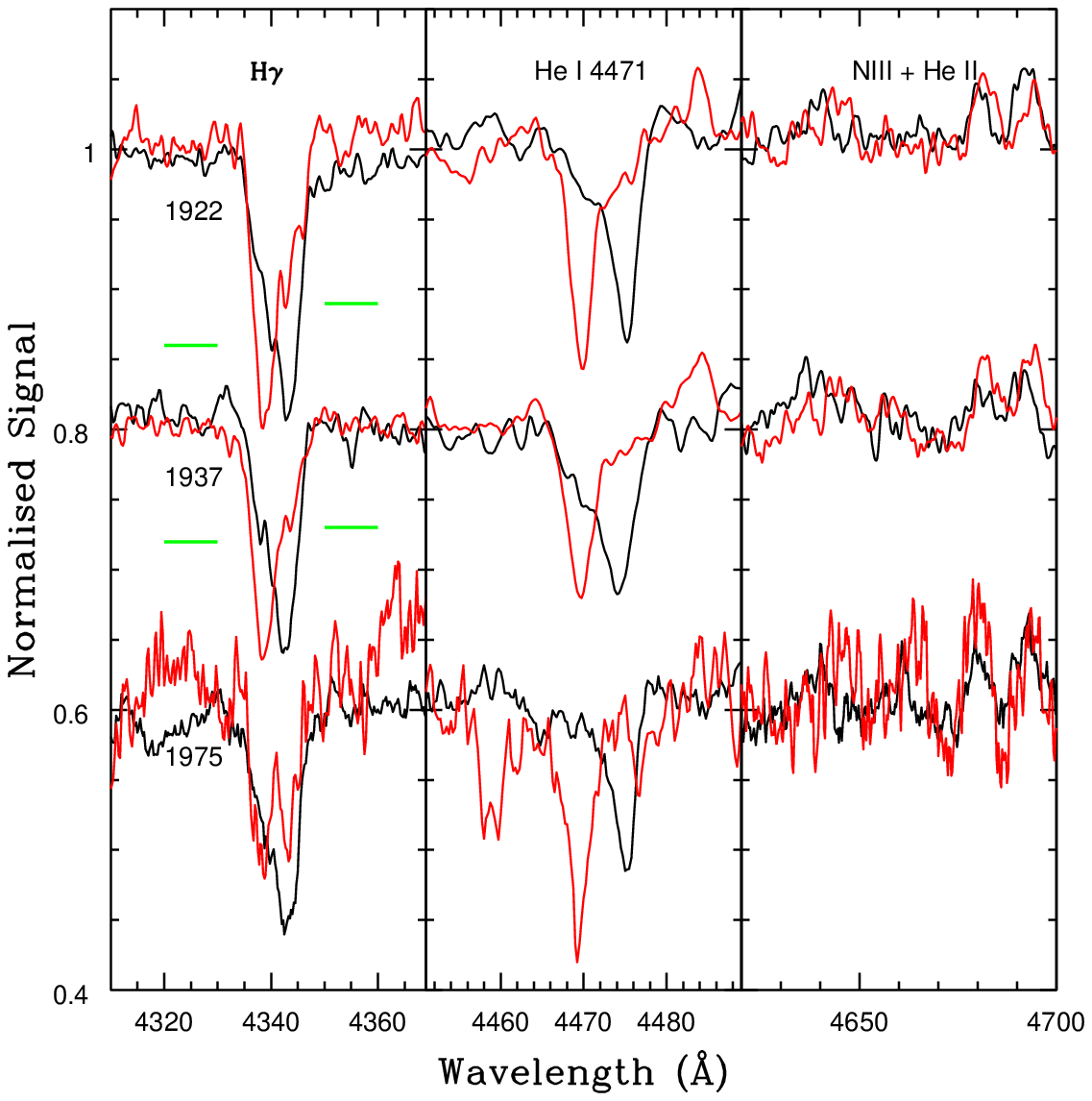}
\caption{Comparing line profiles near phases 0.25 and 0.75. The black 
lines are the mean profiles for orbital phases 0.15 - 0.35, while the red 
lines are for phases 0.65 - 0.85. The spectra 
for phases 0.65 -- 0.85 from 1937 have been scaled by 
a factor of 1.5 to adjust for the difference in line depths seen 
in Figure 7. The line profiles do not mirror each 
other, in the sense that the notch produced 
by the spectrum of the secondary, which is marked with the green 
lines near the 1922 and 1937 H$\gamma$ profiles, 
does not occur at the same depths at the two phases. This is 
most noticeable in the H$\gamma$ profiles of the 1922 campaign spectra. To the 
extent that H$\gamma$ tracks the orbital kinematics of the system, then these 
comparisons indicate that the systemic velocity measured for the 
secondary from the 1922 campaign spectra differs from that obtained 
from the 1937 campaign spectra, and this is confirmed with a cross-correlation 
analysis in Section 6. There is also a tendency for the \ion{N}{3} 
emission lines near 4640\AA\ and for \ion{He}{2} $\lambda 4686$  
emission/absorption to move with wavelength in a manner that is consistent 
with the secondary. The presence of \ion{N}{3} and \ion{He}{2} $\lambda 4686$ 
emission is consistent with an Of supergiant designation for that star.} 
\end{figure}

	The green lines in Figure 9 mark the location of the 
notch from the secondary in the H$\gamma$ profile. 
The location of this notch in the 1922 spectrum 
is three quarters of the way down the H$\gamma$ line between phases 
0.15 and 0.35, but is only half way down the profile at phases 0.65 -- 0.85. 
This asymmetry indicates that the systemic velocity estimated from the 
radial velocity curve of the secondary differs from that of the 
primary. This is consistent with the earlier analyses by \citet{pla1922} and 
\citet{sti1997}, both of whom found that the components have different mean 
space velocities.

	The disagreement between systemic velocities does not occur at 
all epochs. The notch due to the secondary in the 1937 H$\gamma$ profile occurs 
at roughly the same depth near phases 0.25 and 0.75 indicating that the 
systemic velocities obtained from the radial velocities of the primary 
and secondary from those data should agree. Differences in the kinematic 
properties of the 1922 and 1937 campaign spectra are examined further 
in Section 6, where the CCFs of individual spectra near phases 0.25 and 0.75 
are compared, as well as in the Appendix, where radial velocities 
obtained from individual spectra are discussed. 

	What could cause the systemic velocities of the 1922 and 1937 
spectra to differ? If there is an optically thick absorbing medium with 
bulk motions that differ from the orbital motions of the stars, then 
variations in the radial velocities measured from absorption lines may not 
track faithfully the orbital motions of the component stars. We suggest that 
the behaviour of the secondary spectrum in the H and He profiles might 
be affected by an optically thick envelope around that star. If there is 
transient structuring in the envelope, then this could introduce 
variations in its velocity measurements with time. 

	There is evidence of a circumstellar 
envelope around the secondary. \citet{wigandgie1992} argue 
that the wind from the secondary dominates over that from the primary, and 
the wind will produce a circumstellar envelope. Signatures of an 
extended envelope around one or both stars are seen in the DAO spectra. 
The properties of the \ion{N}{3} and \ion{He}{2} $\lambda 4686$ emission lines 
are compared in the middle column of Figures 6, 7, and 8. 
The characteristics of \ion{N}{3} emission near 4630\AA\ in Figures 6 and 
7 change with orbital phase, with the lines being widest near phase 0.75. 
The \ion{N}{3} lines are likely photospheric in origin 
\citep[e.g.][]{bruandmih1971, heaetal2006}, and phase-related variations are 
then indicative of asymmetries in the photospheric properties of one or 
both stars. The relative strengths of the \ion{N}{3} lines also change with 
time with respect to \ion{He}{2} $\lambda 4686$ 
emission, in the sense that these features are weaker near 
phase 0.25 in the 1922 spectra when compared with the 1937 spectra.
Morphological variability in \ion{He}{2} $\lambda 4686$ is seen in O 
supergiants with spectral types that are similar to the stars in PS, 
such as HD192639 \citep[]{rauandvre1998}.

	The emission in the DAO spectra is likely dominated by the secondary 
and its surroundings. The \ion{N}{3}, \ion{He}{2} $\lambda 4686$, and H$\beta$ 
emission in Figure 9 moves in wavelength with orbital phase in the 1922 and 
1937 spectra in a manner that is similar to that 
of the secondary, with the overall change in wavelength 
being comparable to that found for absorption features 
associated with that star. A similar shift is not seen in the 1.2 meter 
1975 spectra of \ion{He}{2} $\lambda 4686$, although the S/N of those spectra 
is lower than that of the 1.8 meter spectra. There is no 
evidence for emission associated with the primary. As line emission 
is associated with the 'f' designation for O stars then these data suggest 
that an 'f' designation should be assigned to the spectral type of 
the secondary, but not the primary. 

	We close this section with a brief discussion of the morphology 
of \ion{He}{2} $\lambda 4686$, and its relation to the properties of the 
secondary.  \citet{venetal2002} model the behaviour of \ion{He}{2} lines in 
a spherically symmetric expanding atmosphere, and link 
the behaviour of \ion{He}{2} $\lambda 4686$ to the properties of the envelope 
and the host star. For models in which the temperature profile peaks at 
intermediate radii or climbs steadily with radius, there is an increased 
tendency for \ion{He}{2} $\lambda 4686$ to show the double-peaked morphology 
seen in the 1.2 meter spectra (1) as log(g) is lowered, and/or (2) 
when the emitting region becomes more extended. In the context of 
these models, a double peaked \ion{He}{2} $\lambda 4686$ morphology 
is consistent with a surface gravity for the secondary that is lower than 
that of main sequence stars. 

	The morphology of the \ion{He}{2} $\lambda 4686$ 
profile is also sensitive to the effective temperature of 
the host star. While the \citet{venetal2002} models 
do not predict a double peaked profile for an effective temperature of 40000K, 
such a profile results for effective temperatures 
of 25000 K. The effective temperature of the secondary in PS is 
$\sim 33000$ K \citep[]{linetal2008}, and so falls midway 
between the temperatures modelled by \citet{venetal2002}. 
The general nature of the \ion{He}{2} $\lambda 4686$ feature in the PS 
star is thus broadly consistent with the \citet{venetal2002} models.

	\citet{puletal2020} examine model line profiles for 
hot stars that assume spherical symmetry among the deeper portions of the 
atmosphere. The models have an approximate resolving power of $10^4$, 
which is higher than that of the 1.8 meter and 1.2 meter spectra. 
The double-peaked morphology that is seen in the 1.2 meter spectra, 
but not the 1.8 meter spectra, is not produced in the models.

	The \citet{puletal2020} model that comes closest to 
matching the effective temperature and log(g) of the components in PS is s6a, 
which produces a modest P Cygni-like profile for \ion{He}{2} $\lambda 
4686$. The absorption and emission components have an amplitude of 
only a few percent of the continuum level. While this does not match the 
appearance of the \ion{He}{2} profile in the spectra, the character of the 
profile is temperature-sensitive, and a discernable absorption component 
disappears in the line profile when the effective 
temperature is increased to 38700 K. Emission is more prominent in this model, 
and amounts to $\sim 10\%$ of the continuum level. The emission line for this 
model is asymmetric in shape, as is seen in some of the 1.8 meter spectra.

\section{COMPARISONS WITH SYNTHETIC SPECTRA}

	Comparisons with synthetic spectra are a means of assessing 
and validating the information content of the digitized spectra. 
Agreement between synthetic spectra and the observations would 
add confidence to our understanding of the observations, and their 
limitations. This is of particular importance for photographic spectra 
such as those recorded with the 1.8 meter telescope
that are prone to a number of factors that have the potential to 
compromise their information content (e.g. Section 4.2).
We have opted not to compare the 1.2 meter spectra with synthetic spectra given 
the modest number of high-quality spectra in that sample. Still, the 
comparisons made in Figure 2 suggests that the conclusions reached here 
should also apply to those spectra.

	Synthetic spectra that cover 4300 -- 4551\AA\ were considered. This 
is near the central wavelength of the observations and contains H$\gamma$ and 
\ion{He}{1} $\lambda 4471$, which are two of the strongest features in the 
blue/visible spectrum of late-type O stars. H$\gamma$ is less susceptible 
to line emission than H$\beta$; while emission is clearly present 
in the wings of the latter, it is much more subdued in the former.
As the goal of the comparisons is not to use the synthetic spectra to 
extract or fine tune system parameters, but to assess if the observed 
spectra -- when considered on average -- agree with what 
might be expected from spectra recorded with modern detectors, then 
a forward modelling approach is adopted.

\subsection{Models}

	The synthetic spectra are based on the solar chemical 
composition OSTAR2002 models discussed by \citet{lanandhub2003}
that were generated with TLUSTY \citep[]{hub1988, hubandlan1995}.
The baseline parameters used to construct the synthetic 
spectra of PS are the effective temperatures (T$_{eff}$), 
surface gravities (log(g)), fractional contributions 
to the light near $0.45\mu$m (f$_B$), observed rotational velocities 
($vsini$), and the orbital velocity semiamplitudes (K) 
of both components. The value adopted for each of these to 
construct the synthetic spectra is shown in Table 3. The entries in 
this table are taken from Table 3 of \citet{linetal2008}, with the $vsini$ 
values being the means of the entries in his table. The quantities 
in Table 3 generally agree with those found in other studies, and 
experimentation revealed that tweaking the effective temperatures and surface 
gavities by small amounts does not greatly alter the synthesized spectra. 

\begin{center}
\begin{deluxetable}{lccl}
\tablecaption{Parameters Adopted for Synthetic Spectra}
\tablehead{Parameter\tablenotemark{a} & Star 1 & Star 2 & Notes \\}
\startdata
T$_{eff}$ & 33500 & 33000 & \\
log(g) & 3.5 & 3.5 & \\
$vsini$ & 66 & 273 & (km/sec) \\
K & 202.2 & 192.4 & (km/sec) \\
f$_B$ & 0.65 & 0.35 & \\
\enddata
\tablenotetext{a}{All quantities are from \citet{linetal2008}}
\end{deluxetable}
\end{center}

	The component stars are the only sources of light that are 
considered in the models. To be sure, there 
are other sources of light in interacting binary 
systems, such as emission from disks and outflows. The component stars 
may also have spots that will alter their spectrophotometric properties. 
However, as will be demonstrated later in this section, the 
basic model adopted here reproduces many of the 
large-scale spectroscopic characteristics of the system. 

	A spectrum was generated for each component, 
and the results were smoothed according to their adopted rotational 
velocities. The synthetic spectra of the components were 
scaled according to their light contributions, shifted according to the 
velocity that is appropriate for the orbital phase, and then added 
together. A model spectrum of the system was constructed for 
phases near 0.25 and 0.75, as well as near phase 0.0. 

\subsection{Comparisons with the Observations}

	The results are compared with the means of 
spectra within the different phase intervals in 
Figure 10. The spectra in this figure are the means of all 
the 1.8 meter spectra within each phase interval that were recorded between 
1922 and 1938 that are not of poor quality. Not restricting the means to 
spectra recorded during the campaigns resulted in a more uniform 
sampling of phase coverage within each interval.

\begin{figure}
\figurenum{10}
\epsscale{1.0}
\plotone{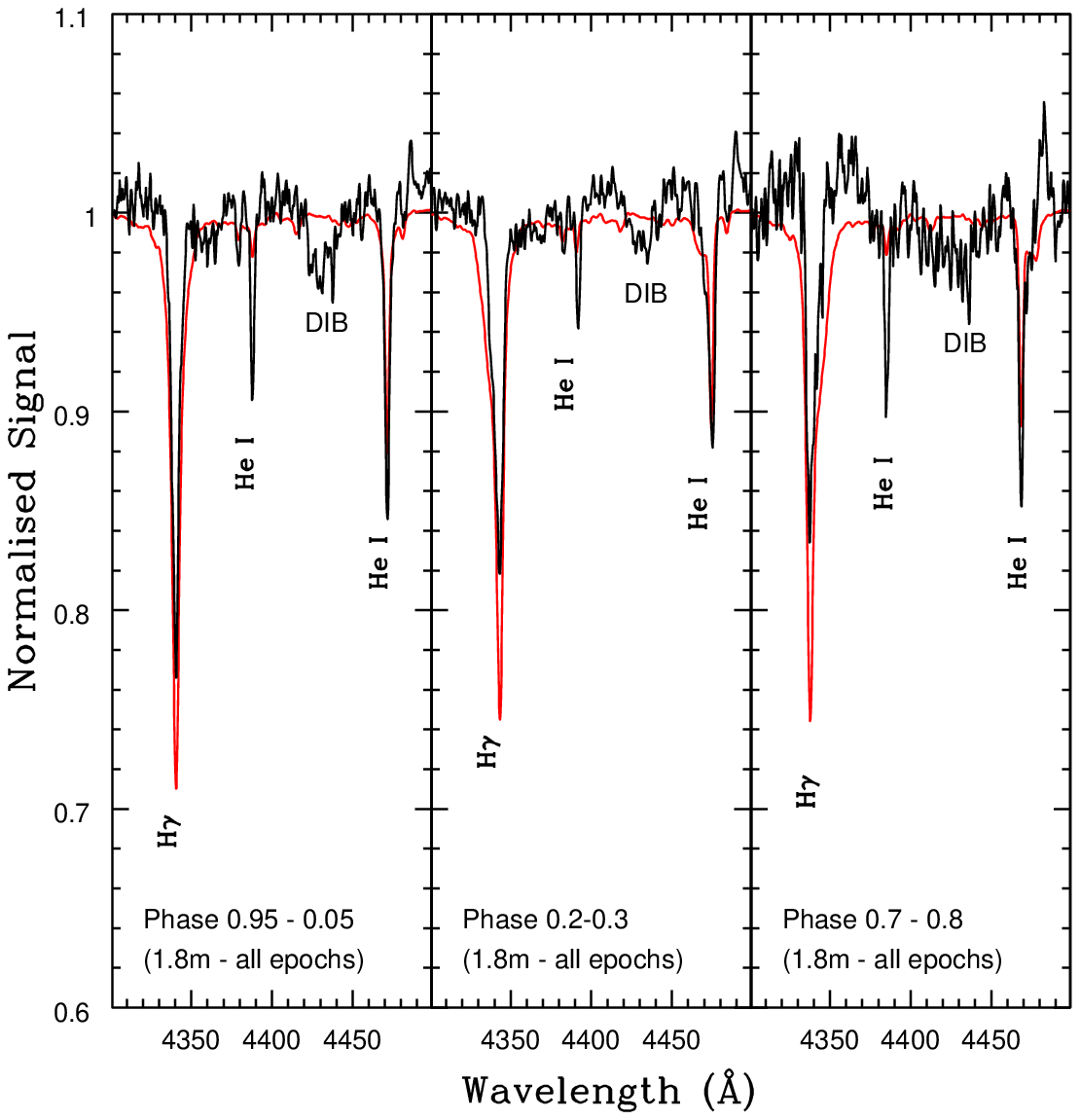}
\caption{Comparisons with synthetic spectra. The spectra of 
PS are the means of many of the 1.8 meter spectra recorded between 
1922 and 1938 that fall within each phase interval. Spectra 7302 and 7006 were 
excluded because of poor S/Ns (see text), as were spectra 26304 
and 26329 from the 1937 campaign. H$\gamma$ in the synthetic spectrum is 
consistently deeper than that observed. There is better agreement with 
\ion{He}{1} $\lambda 4471$ at phases 0.2 - 0.3 and 0.7 -- 0.8. The models 
consistently underestimate the depth of 
\ion{He}{1} $\lambda 4388$. A comparison of the 
synthetic and observed line profiles near phases 
0.25 and 0.75, including the location of the notch produced by the 
secondary in the profile, suggests that the overall contribution of 
the secondary to the total light from the system is more-or-less 
reproduced in the synthetic spectrum. However, that the synthetic 
spectra are broader than the upper portions of H$\gamma$ and 
\ion{He}{1} $\lambda 4471$ near phases 0.25 and 0.75 is consistent with 
the orbital velocity of the secondary having been overestimated.} 
\end{figure} 

	There is an even mix of spectra from the 1920s and 1930s in the 
0.95 -- 0.05 and 0.2 -- 0.3 phase intervals. Spectrum 7006 was not included 
when computing the mean spectrum between phases 0.2 and 0.3 because it 
has shallow absorption lines (e.g. Figure 6). The mean spectrum in 
the 0.7 -- 0.8 phase interval is dominated by 1937 campaign spectra, 
as only two 1922 campaign spectra fall within this phase interval 
and were used when taking the mean. Spectrum 7302 has a poor S/N and was 
not included. 

	There is mixed success matching the observations. The synthetic 
spectra overestimate the depth of H$\gamma$ at all three phases, with the 
agreement between the modelled and observed depth of H$\gamma$ 
closest between phases 0.95 and 0.05. However, there is approximate 
agreement between the observed and predicted depths of 
\ion{He}{1} $\lambda 4471$ at all three phases. 
The synthetic spectra also more-or-less reproduce the skewed line shapes at 
the orbital quadrature points, although the line widths are not reproduced, 
in the sense that the model spectra are broader.
The differences between the synthetic spectra and observations 
are in the sense that the amplitude of the orbital velocity of the secondary 
appears to have been overestimated when constructing the models. The 
issue of the orbital velocity of the secondary is discussed at greater 
length in Section 6 and the Appendix.

	\ion{He}{1} $\lambda 4388$ stands out as it is much deeper 
than in the synthetic spectra. This line is a singlet 
transition, and its strength in models is susceptible to uncertainties 
in the properties of the spectrum near \ion{He}{1} $\lambda 584$ 
\citep[][]{najetal2006}. This being said, the inability to reproduce the 
depth of \ion{He}{1} $\lambda 4388$ is not common 
to all TLUSTY models. \citet{gonandlei1999} generate a grid 
of spectra that use simplified assumptions for opacity and are restricted 
to predicting the strengths of only H and He lines. The depth of 
\ion{He}{1} $\lambda 4388$ in their models more closely matches that 
observed in PS. The equivalent widths of H$\gamma$ and \ion{He}{1} 
$\lambda 4471$ predicted by the \cite{gonandlei1999} 
models match very well those in the synthetic spectra 
generated from the \cite{lanandhub2003} models.

	In Figure 1 it was shown that there are some differences between the 
composite 1922 and 1937 campaign spectra, and this leads us to 
examine if the synthetic spectra match one campaign dataset better 
than the other. The synthetic spectra are compared with mean spectra from 
the 1922 and 1937 campaigns in Figure 11. The phase 
interval 0.7 -- 0.8 is not included given that there is only one 
spectrum in the 1922 campaign (plate 7301) that has a useable S/N. 

\begin{figure}
\figurenum{11}
\epsscale{1.0}
\plotone{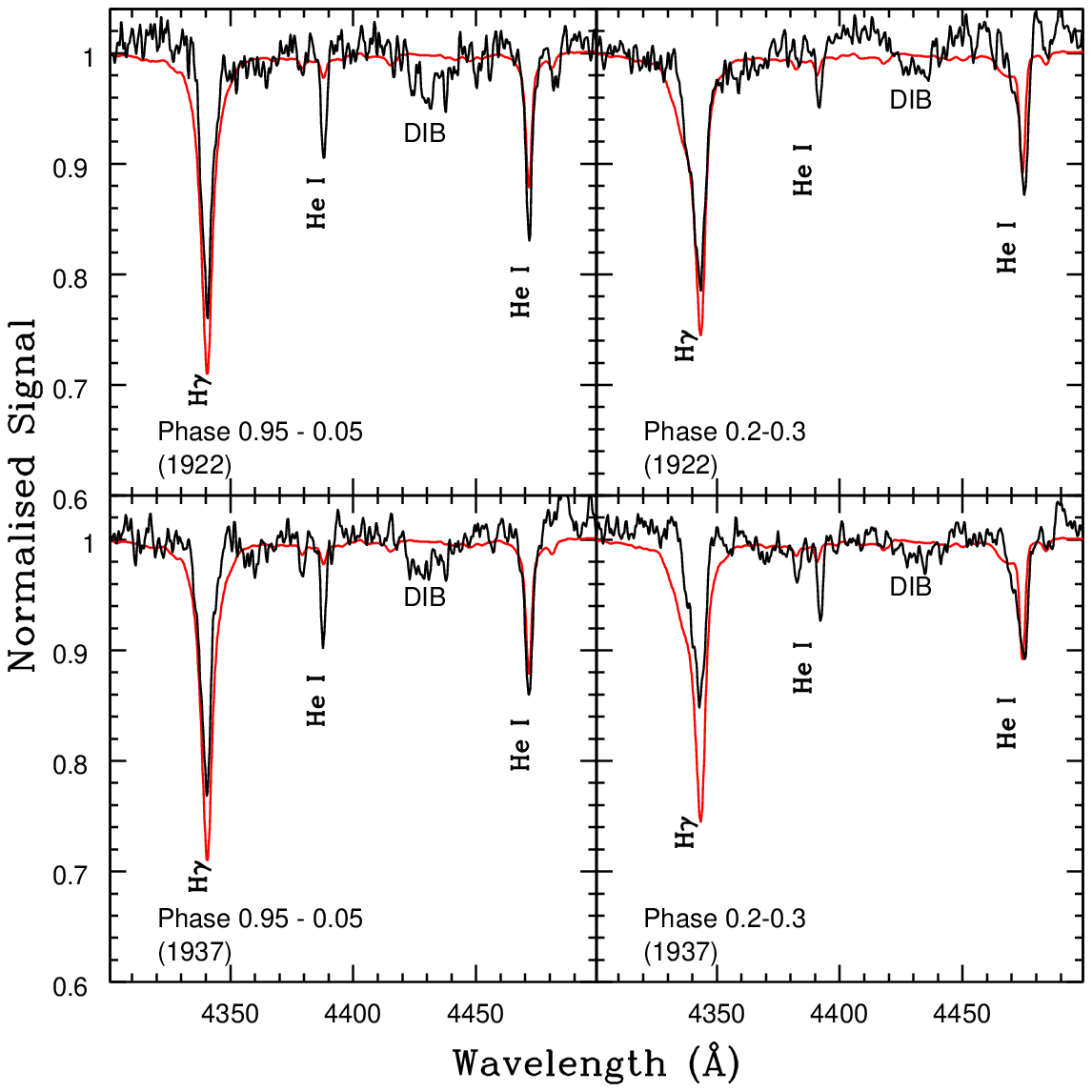}
\caption{Comparing campaign and synthetic spectra. 
The spectra of PS are the means of the 1.8 meter spectra 
from the 1922 (top panel) and 1937 campaigns (bottom row). 
The phase interval 0.7 -- 0.8 is not included because 
there is only one useable spectrum that samples this interval in the 1922 
dataset. H$\gamma$ is consistently larger in the 
synthetic spectra than in either group of campaign spectra, 
with the difference being greatest for the 1937 spectra
between phases 0.2 and 0.3. In contrast, the depth of 
\ion{He}{1} $\lambda 4471$ is more-or-less reproduced by the models. 
It is argued in the text that the comparatively shallow nature of 
H$\gamma$ near phase 0.25 in the 1937 spectrum was a transient phenomenon.} 
\end{figure}

	It was shown earlier that the models did not match the depth 
of H$\gamma$ in the composite spectra, and this 
disagreement is greatest in the 1937 campaign spectrum 
for phases 0.2 -- 0.3. The comparisons made in Figure 3 
indicate that this is not due to only one spectrum; all three 
1937 campaign spectra that were averaged together for this 
phase interval have similar H$\gamma$ depths. These three spectra were recorded 
over the course of four consecutive orbital cycles.
The synthetic spectra are in reasonable agreement with the width and 
depth of \ion{He}{1} $\lambda 4471$ in most 
of the spectra in Figure 11. We thus conclude that H$\gamma$ 
changed strength with orbital phase during the 1937 campaign. 
Moreover, the depth of H$\gamma$ near phase 0 in the 1937 spectrum is 
similar to that in the 1922 spectrum at the same phase, and 
so it is likely that the comparatively shallow nature of 
H$\gamma$ between phases 0.2 and 0.3 during the 1937 campaign was a transient 
phenomenon that was not present during the 1922 campaign. 

	If the depths of features in the PS spectra are taken at face value 
then a better match to the depth of H$\gamma$ may be found if the effective 
temperature of the primary in the models were to be increased and/or if the 
surface gravity were changed at some phases. This could occur if the 
temperature distribution across the face of the primary was 
not uniform, perhaps due to a spot or a reflection effect from a hot area 
associated with the secondary and/or its surroundings. The TLUSTY-based model 
H and He line strengths discussed by \cite{gonandlei1999} 
provide a convenient means of assessing the dependence of H 
and He line strengths on parameters such as effective temperature and 
surface gravity.

	The \cite{gonandlei1999} models suggest that if 
the side of the primary that faced the observer at 
phase 0.25 during the 1937 campaign had an effective temperature 
that was $\sim 5000$K hotter than the rest of the star 
then this could largely explain the relative depths of H$\gamma$ and 
\ion{He}{1} $\lambda 4471$. However, models with a main sequence surface 
gravity are then required for the primary if its spectrum is to match 
the observed H$\gamma$/\ion{He}{1} $\lambda 4471$ ratio, and 
this is contrary to the luminosity class 
assigned to this star by most studies. Moreover, the ratio of \ion{He}{2} 
$\lambda 4541$ to \ion{He}{1} $\lambda 4471$ is a 
gauge of effective temperature, and this line ratio 
among spectra near phase 0.25 in the 1937 campaign spectrum agrees 
well with that measured from the 1922 campaign spectrum -- the relative depths 
of these lines are thus consistent with the effective temperature not 
changing with time.

	The comparisons with the models thus lead us to suspect 
that the relatively shallow depth of H$\gamma$ in the 1937 
campaign spectra near phase 0.25 cannot be due to 
surface temperature variations. While not explored here, perhaps the 
variations in the depth of H$\gamma$ may be due to abundance spots, such 
as those seen in Ap and Bp stars. In any event, the possibility that the depth 
of H$\gamma$ is affected by the response of the photograpic plate seems 
unlikely. This is because the spectra recorded in 1937 near 
this phase over different nights have similar H$\gamma$ depths. A similarity 
in bogus line depths would then be due to either a 
statistical fluke or a systematic issue in the plates that 
were used for the observations near phase 0.25, but not at other phases,
which is highly improbable. 

\section{CROSS-CORRELATION FUNCTIONS AND THE MASS RATIO}

	Lines with different excitation conditions at blue wavelengths 
in the PS spectrum yield similar velocities \citep[]{bagandbar1996}, 
and so the information content in the PS spectra can be multiplexed by 
using a cross-correlation procedure to measure velocities. 
In addition to radial velocities, CCFs 
also contain information about the overall strengths of features 
in the spectra of the components. The CCFs discussed here were generated 
with the FXCOR task in IRAF, which is based on a program 
described by \citet{tonanddav1979}. 

	The CCFs discussed in this section were 
generated in the 4300 -- 4600\AA\ wavelength interval, which samples the peak 
optical response of the 1.8 meter spectra. The largest single contributors 
to the CCFs are then H$\gamma$ and \ion{He}{1} $\lambda 4471$, 
although weaker features also contribute to the CCFs. 
Extending the coverage to longer wavelengths to include H$\beta$ produces 
slightly noisier CCFs, although the location of the peaks associated with 
the components do not change markedly. Velocities 
obtained from the 1.8 meter spectra using the 
4300 -- 4600\AA\ range and those that use the full wavelength coverage of the 
spectra are compared in the Appendix.

	The reference spectrum used to obtain the CCFs is a combination 
of} H, He, and N lines with depths that are taken from the median 1922 
campaign spectrum in Figure 1. To the extent that the spectrum of the secondary 
is suppressed when spectra that are aligned on the 
centers of strong lines in the primary are combined, then 
the template is a proxy for the spectrum of the primary. 
Absorption lines in the template are $\delta$ functions 
that extend down from a flat continuum, and the template is compared with 
1922 campaign spectra near orbital phases 0.25 and 
0.75 in Figure 12. The use of an unbroadened spectrum as a reference 
template more cleanly separates the peaks in the CCF that 
form from lines in the spectra of the two components, 
which can blend together if real stellar spectra are used as 
a reference, such as in Figures 5 and 6 of \citet{bagetal1999}.
The use of an unbroadened template for velocity measurements was first
suggested by \citet{furandfur1990}.

\begin{figure}
\figurenum{12}
\epsscale{1.0}
\plotone{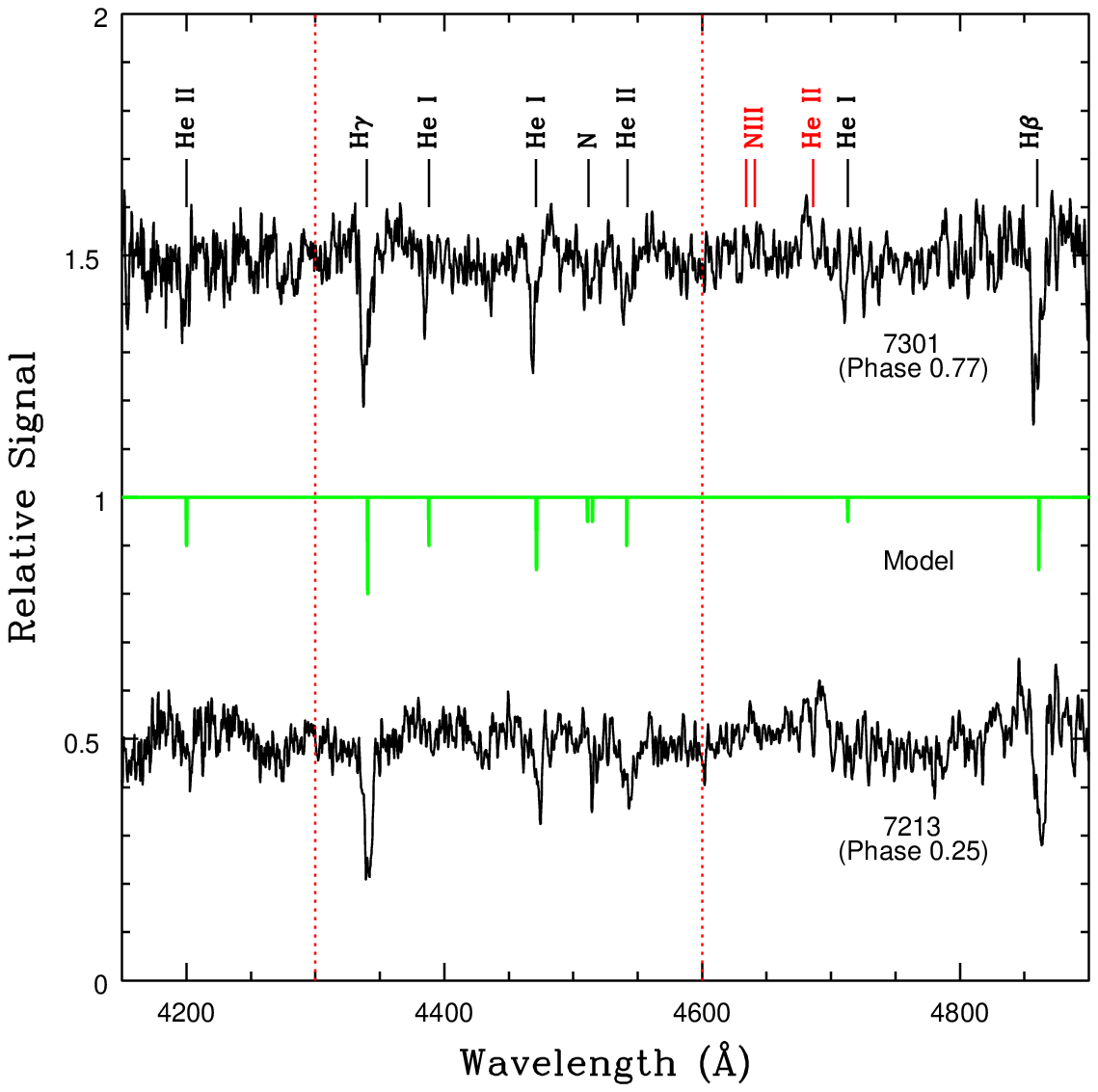}
\caption{Spectra from the 1922 campaign that sample orbital phases 
near 0.25 and 0.75 are compared with the CCF template, shown in green. The 
template consists of unbroadened lines that correspond to H, He, and N 
transitions, with depths based on those of the actual lines in the 
median 1922 campaign spectrum. The dotted lines mark the boundaries of the 
wavelength interval used to generate the CCFs.}
\end{figure}

	Emphasis is placed on orbital phases 0.25 and 0.75, where the velocity 
difference between the stars is greatest, thereby yielding the 
half amplitudes of the stellar motions (K$_1$ and K$_2$). 
There are multiple spectra between phases 0.2 and 0.3 in the 1922 and 1937 
campaign datasets. However, while there are five spectra between 
phases 0.7 and 0.8 in the 1937 campaign dataset, there is only one 
spectrum (7301) from the 1922 campaign in this phase interval 
that is suitable for velocity measurements; spectrum 7302 
has a very poor S/N, and produces a problematic CCF. 

	Two spectra from the 1.2 meter observations are also considered: 8820 
at phase 0.2, and 9545 at phase 0.73. While the former is not part of the 1975 
campaign, it is the only useable 1.2 meter spectrum that samples a phase close 
to 0.25. The 1.2 meter spectra are of interest when interpreting the 
CCFs as they were recorded with an image slicer, 
and so are less susceptible to variations in spectral resolution due to 
seeing variations. The 1.2 meter spectra are intrinsically noisier 
than the 1.8 meter spectra, and so they were smoothed with a 0.25\AA\ 
Gaussian to enhance the S/N before generating the CCFs. 

	CCFs that sample orbital phases near 
0.25 and 0.75 are shown in Figures 13 (1.8 meter)
and 14 (1.2 meter). Distinct peaks that are associated with the 
primary and secondary are clearly visible in Figure 13. While not as well 
defined as in Figure 13, features that can be identified with the 
secondary are also apparent in Figure 14.

	The CCFs differ in appearance 
from those obtained by \citet{sti1997}, in the sense that 
the difference in heights between the peaks associated 
with the primary and secondary are much larger than found 
by \citet{sti1997}. This difference is perhaps not surprising as 
the Stickland CCFs were constructed from UV spectra. The difference 
in peak heights reflects the relative light contributions from the 
two stars, which are more closely matched in the UV than near 4400\AA\ .

\begin{figure}
\figurenum{13}
\epsscale{1.0}
\plotone{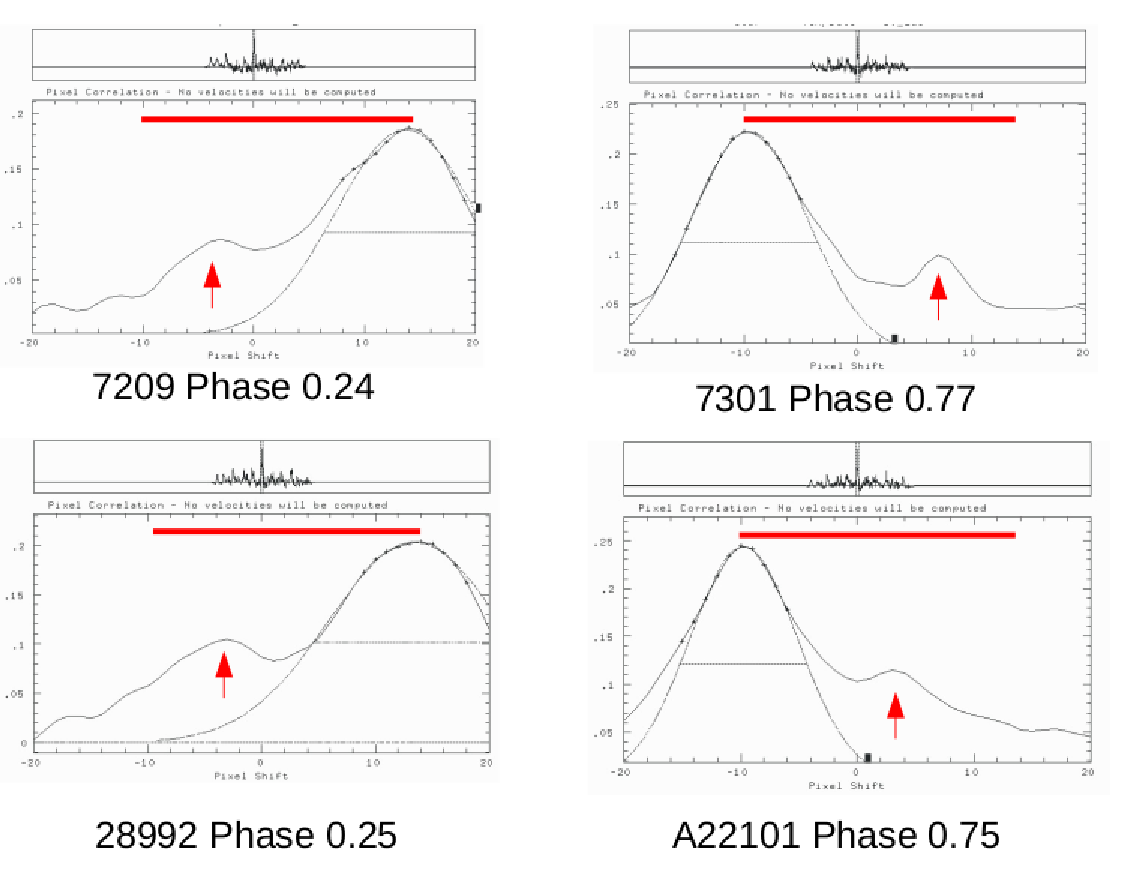}
\caption{CCFs of 1.8 meter spectra near orbital phases 
0.25 and 0.75 from the 1922 (top row) and 1937 (bottom 
row) campaigns. The spectra have been cross-correlated 
with the reference spectrum between wavelengths 4300 and 4600\AA\ using 
the IRAF task FXCOR. The dominant peak in the CCFs is formed by lines in the 
spectrum of the primary, while a lower amplitude peak due to the secondary is 
marked with a red arrow. The thick red line has a length of 
400 km/sec, which is the approximate full amplitude of the 
radial velocity variation of the primary. The separation between the 
peaks due to the primary and secondary varies, corresponding to 
$\sim 100 - 190$ km/sec, which is smaller than 
the difference in component velocities found in most 
other studies. That the radial velocity amplitude of 
the secondary is smaller than that of the primary is consistent with it being 
the more massive star. The dashed line is a Gaussian fit to the 
peak from the primary. The width of the peak in the CCFs 
formed by the primary near phase 0.25 is greater than that near phase 0.75, 
and this is also seen in the CCFs of the 1.2 meter spectra in Figure 14.}
\end{figure}

\begin{figure}
\figurenum{14}
\epsscale{1.0}
\plotone{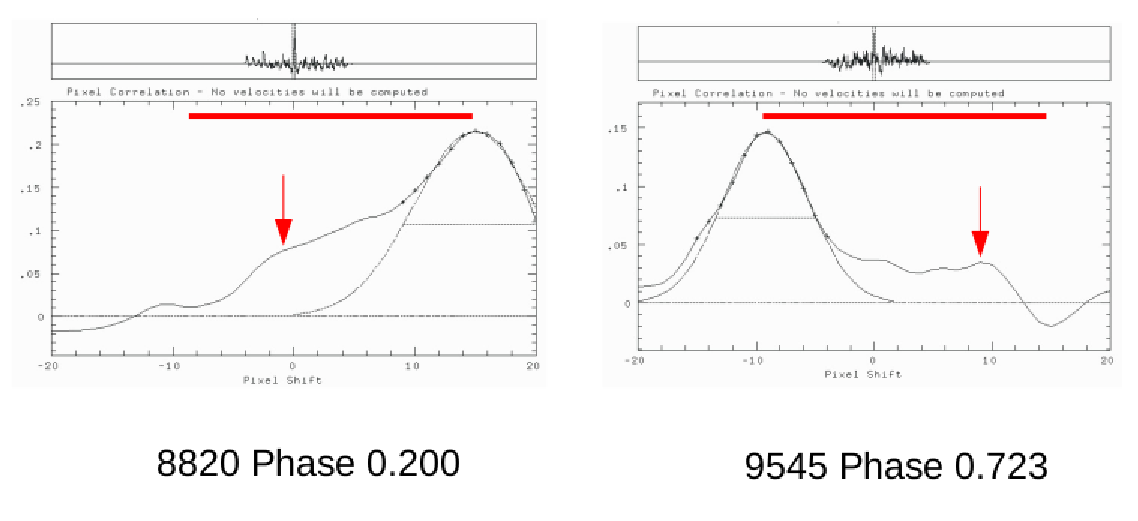}
\caption{Same as Figure 13, but showing CCFs constructed from 1.2 meter 
spectra. The peak due to the secondary is less well-defined than in Figure 13, 
reflecting the lower S/N of the 1.2 meter spectra. Still, 
the velocity amplitude based on these CCFs is consistent with 
that obtained from the 1922 campaign spectra. As in Figure 13, the 
peaks are broader near phase 0.25 than near phase 0.75. As the 1.2 
meter spectra were recorded with an image slicer then it is unlikely 
that this broadening is a consequence of seeing variations.}
\end{figure}

	The separation between the peaks that correspond to the 
primary and secondary in Figures 13 and 14 are smaller 
than found by \citet{sti1997}, amounting to 
100 -- 190 km/sec as opposed to 300 km/sec from the Stickland CCFs 
of SWP 4819 and SWP 13924 in his Figure 4. In the Appendix it is argued that 
this is likely due in large part to the wavelength resolution of the 
DAO spectra. Still, the difference in velocity between the two 
components obtained from spectrum SWP 4819 in Figure 4 of \citet{sti1997} is 
smaller than that in spectrum SWP 13924. This, along with other factors, 
lead Stickland to conclude that the velocities of the secondary 'make no sense 
dynamically'. 

	The peak associated with the secondary near phase 0.75 in 
Figure 13 changes location with time, in the sense that the offset between the 
signatures of the primary and secondary in the CCF of spectrum 7301 
from the 1922 campaign is larger than that in spectrum A22101 from the 
1937 campaign. It should be recalled that line shapes in the 1922 and 1937 
campaign spectra at phase 0.25 differ because of the 
location of the notch associated with the secondary in each 
profile (Section 4). The difference in line profiles is consistent with the 
behaviour of the secondary feature in Figure 13.

	While establishing a velocity amplitude for the secondary is 
complicated by the variations in the location of the peak attributed to 
that star, it is clear that the amplitude of the orbital 
velocity variations of the secondary is smaller than that of the primary. 
The smaller amplitude of the radial velocity curve 
of the secondary is consistent with it being more massive than 
the primary. If mass transfer has occured then PS is a post-Algol binary.

	Based on the velocity measurements discussed in the Appendix, 
we estimate a velocity half amplitude of 90 km/sec for the 
secondary, and a ratio of amplitudes of $0.5 \pm 0.1$, in the sense 
primary over secondary. The mass ratio is then $2.0 \pm 0.4$. This is 
substantially larger than most previous estimates, and yields lower 
component masses than found in previous studies. Adopting the \cite{sti1997} 
mass function, then with a mass ratio of 2 $M{_1}sin(i)^3 = 14.2 \pm 1.2$ 
and $M_{2}sin(i)^s = 28.3 \pm 2.7$ M$_{\odot}$.

	The orbital inclination is uncertain as the 
system is not eclipsing. \citet{bagetal1992} argue for a value between 69.3 
and 72.7 degrees based on the radius of the components, while 
\citet{rudandher1978} estimates an inclination of $71 \pm 9$ degrees from 
polarimetry. Adopting $i = 71^o$ yields masses of 17 and 34 M$_{\odot}$, 
while if i = 62$^o$ based on the lower limit from \citet{rudandher1978} 
then the masses are 21 and 42 M$_{\odot}$.

	Given the substantial uncertainties in 
the properties of the secondary and the difficulties extracting 
its spectroscopic signatures from these moderate dispersion spectra
then the mass ratio and mass estimates should be viewed 
with caution. Arguably the most robust conclusion 
that can be drawn from the CCFs in Figures 13 and 14 is that the secondary is 
more massive than the primary. Determining an improved mass ratio 
will require a better understanding of what is causing the velocity 
measurements that are attributed to the secondary to drift with time.

	In Section 4.3 it was suggested that the system line profiles 
might be explained if there is a component in the spectrum that is 
related to the secondary but that also has velocities that 
do not track faithfully the orbital motion of that star. 
If the wind from the secondary is the dominant contributor to 
material being fed into the intracomponent medium \citep[]{wigandgie1992}, 
then we suggest that the velocities obtained from the 
absorption spectrum that is associated with the secondary may be skewed 
by transient optically thick material in the outflow from that star. 
Even small-scale perturbations in the winds of evolved O stars are expected 
to induce clumping \citep[e.g.][and references therein]{hawetal2021}, 
and we suggest that it is the transient motion of such clumps that might 
affect the radial velocity measurements that are attributed to the 
secondary.

\section{DISCUSSION \& SUMMARY}

	We have discussed photographic spectra of PS that were recorded 
with the 1.2 and 1.8 meter telescopes at the DAO. 
The spectra sample the wavelength region $0.39 - 0.49\mu$m 
(1.8 meter) and $0.35 - 0.55\mu$m (1.2 meter) with wavelength 
resolutions of a few thousand. The spectra were 
digitized with a flat-bed scanner, and the results were processesed according 
to the procedures described by \citet{dav2024}.

	The spectra were recorded over a 5 decade time span from the 
early 1920s to the mid 1970s, and provide insights into the evolution of 
PS during much of the past century, prior to the widespread use of digital 
detectors. Many of the 1.8 meter spectra were obtained 
during campaigns in the 1922 and 1937 observing seasons, and so 
provide snapshots of the PS system at those times. 
The 1922 campaign spectra were the basis of the 
discovery paper by \citet{pla1922}, and are among the oldest extant spectra of 
this system. To the best of our knowledge, the spectra recorded in the 1930s 
have not been discussed in the published literature.

	The S/N of the data used here is admittedly low, and can 
easily be surpassed with even short duration exposures with modern detectors. 
Still, spectra of this nature are the only means available to study the 
properties of stars prior to circa 1980. As of late 2024, the 
PS system has undergone over 2600 orbital cycles since the first 
1.8 meter spectrum was recorded. The DAO spectra thus provide an 
anchor for examining the long term evolution of the PS system.

	A complication when examining spectra of this vintage is that 
they do not form a homogeneous dataset. The quality of the 
digitized DAO spectra range from very poor to very good, and this 
is due in part to environmental factors, such as cloud cover, seeing 
variations, etc. There are also spectra where there is evidence of 
issues with the instrumentation, such as the mis-alignment 
of optical components, as well as with the handling and development of the 
photographic plates. Plates that will likely be problematic for 
future studies have been identified. 

\subsection{Line Strengths and Velocity Variations}

	Information has been extracted from the digitized spectra that is 
supplemental to that found in earlier studies where the plates were 
examined in their undigitized form. Comparisons of spectra recorded 
within a campaign and between campaigns have been made. While the response 
characteristics of photographic materials is certainly a limiting factor, the 
overall agreement between spectra within a campaign and between campaigns 
after they have been combined to suppress the effect of environmental factors 
on the observations suggests that {\it on average} the spectra have similar 
response characteristics in the intensity range covered by the spectra. 

	Variations in the strengths of some lines, most noteably 
H$\gamma$, have been found, and evidence has been presented 
that these are intrinsic to the system spectrum.
Variations in the strengths of lines in the spectrum of an SB2 that 
contain early-type stars are not rare, and a subset of these 
variations have been termed the Struve-Sahade (S-S) effect. 
Line strength variations due to the S-S effect 
are typically attributed to the spectrum of the secondary, in the sense that 
the lines are weaker when the star is receding and stronger when it is 
approaching. The physical cause of the S-S effect remains a matter of 
discussion \citep[e.g.][]{linetal2007, paletal2013}. However, it 
is not clear if a classic S-S effect has been detected in the PS spectrum 
\citep[e.g.][]{sti1997, bagetal1999, linetal2008}. The variations we see 
in the strength of H$\gamma$ are associated with the primary.

	Prominent line emission is seen in the digitized spectra, as expected 
given the 'f' designation that has been assigned to the system spectral type. 
\citet{linetal2008} suggest that \ion{He}{2} $\lambda 4686$ and H$\alpha$ 
emission forms in a wind around the secondary star that is flattened along 
the orbital plane. In fact, the wavelength motion of \ion{He}{2} 
$\lambda 4686$ in the digitized spectra is consistent with it being 
associated with the secondary. This holds also for \ion{N}{3} 
emission and emission seen in the shoulders of H$\beta$. This 
is consistent with the secondary having the dominant wind. The association 
of the \ion{N}{3} emission with the secondary suggests that its spectral type 
should be assigned an 'f' designation.

	The digitized spectra have been compared with synthetic 
spectra that are based on system properties summarized by \citet{linetal2008}. 
Groups of spectra that were averaged to suppress 
environmental and instrumental effects are in broad agreement with 
the synthetic spectra, even though no effort has been made to tune 
the system parameters to achieve an optimum match with the observations. 
The secondary produces features in the observations 
that have counterparts in the synthetic spectra, 
and the locations of these signatures in the system line profiles 
are consistent with previous estimates of the light contribution 
made by the secondary to the total system light output.

	Noticeable departures from the synthetic spectra are seen for 
H$\gamma$ at some epochs as well as \ion{He}{1} $\lambda 4388$. The latter 
is an indicator of luminosity class \citep[e.g.][]{mar2018}, although 
the disagreement between the predicted and observed depths 
of this feature may be due to deficiencies in the extent of 
line blanketing near 584\AA\ in the models \citep[][]{najetal2006}.
The variations in the depth of H$\gamma$, if intrinsic 
to the primary, can not be explained by varying the effective temperature 
of that star in a manner that is consistent with its other properties. 

	While it is difficult to judge the strengths of individual features 
from the secondary in many of the individual DAO spectra, CCFs provide 
a blunt means of judging the composite strength of 
features in the spectra of the components. The relative amplitude of 
the peak attributed to the secondary in the CCFs shown in Figures 13 and 
14 does not appear to vary when compared with that of the primary, suggesting 
that the relative strengths of lines in the spectra of the components 
near the quadrature points remain constant. This is 
indirect evidence against a S-S effect in the spectrum of PS.

	The morphologies of the CCFs near phases 0.25 and 0.75 
obtained from the DAO spectra differ from those constructed from 
UV spectra \citep[e.g.][]{sti1997}, in the sense that the peak attributed to 
the secondary in Figures 13 and 14 has a smaller amplitude with respect to that 
of the primary than is seen in CCFs constructed from 
UV spectra. This is largely due to the difference in 
wavelengths, as the relative contributions made by the components 
to the total system light at a given wavelength depends on factors such as 
the effective temperature and the properties of circumstellar and intrasystem 
material. The template used to construct the CCF from the DAO spectra also 
differs from that employed by \citet{sti1997}, who used 
the spectrum of the O8 star HD99486.

	A more traditional use for CCFs is to measure the motions of 
stars. Evidence was presented in Section 6 that the 
velocities measured from the peak in the CCFs that 
is attributed to the secondary varies with time 
in a manner that suggests they may not faithfully track its orbital motion, 
as noted previouly by \citet{sti1997}. This is clearly seen in the DAO 
spectra, where the peak in the CCF that is attributed to the secondary 
at phase 0.75 changes position with time. 

	Velocity measurements for all DAO 1.8 meter 
spectra near the orbital quadrature points are presented in the 
Appendix, and these indicate that the amplitude of the secondary velocity curve 
did not change between 1922 and 1937. The amplitude of the velocity 
curve of the secondary measured from CCFs is much smaller than that of 
the primary, and is consistent with a mass ratio $\sim 2$. This is markedly 
larger than most previous estimates (but see also \citet{gruetal2022}).

	Previous studies have found inherent uncertainties in the 
spectrum of the secondary that will complicate attempts to find a mass 
ratio. For example, there is considerable scatter in the 
velocities measured for the secondary by \citet{linetal2008}, 
and they attribute this to factors such as line intensity variations. 
These uncertainties can have a sizeable influence on the estimated mass ratio. 
Inspection of Figure 2 of \citet{linetal2008} demonstrates that the 
orbital velocities of the secondary near phases 0.25 
and 0.75 are not well-defined. A velocity curve that were to pass through 
the mean of the points near these phases in \citet{linetal2008} Figure 2, 
as opposed to the peak values, would have a half-amplitude 
that is lower than that adopted by Linder et al. by $\sim 
40$ km/sec. This would change the mass ratio obtained from the Linder 
et al. spectra from 1.05 to 1.3, placing the mass ratio in better 
agreement with that obtained from UV spectra by \citet{bagetal1992}. 
More recently, \citet{gruetal2022} present evidence 
that the secondary rotates with a period of 1.2 
days, and find no evidence of orbital motion in its Stokes V profiles. 
This argues for a lower amplitude radial velocity variation for the 
secondary than has previously been seen, further bringing into question 
the orbital properties of the secondary found in past studies. 

	One possible explanation for the kinematic characteristics 
of the secondary is that the absorption lines attributed to that star may 
reflect transient activity in its circumstellar environment. Modest 
variations in the behaviour of \ion{He}{2} $\lambda 4686$ (e.g. Figure 9) lend 
some support to this argument. Variations in \ion{He}{2} lines are seen in the 
spectra of other evolved O stars, and these have been attributed 
to large-scale inhomogeneities \citep[e.g.][]{rauandvre1998} or 
rotational modulation in the wind \citep[e.g.][]{rauetal2001}.

	The width of the peak in the CCF that is formed by the primary 
changes with orbital phase, and this is seen in observations that span a 
range of epochs and instrumentation. Perhaps most significantly, this 
phase-dependent broadening is evident in CCFs constructed from 
1.2 meter spectra, which were recorded through an image slicer, rather 
than a slit. The variation in the width of the CCF is then
not due to the way in which the entrance to the 
spectrograph was illuminated, as might occur due to seeing variations 
if a slit were in place, but is intrinsic to the spectrum. 

	The change in the width of the primary peak in the CCF 
suggests that the spectroscopic properties of that star are not 
symmetric in the orbital plane, and previous evidence for this is seen 
in the behaviour of H$\gamma$ in the 1937 campaign spectra. There 
are only modest variations in the system light output with orbital phase, 
and so any physical mechanism that drives changes 
in the line profiles of the primary, that in turn sets the 
width of the primary peak in the CCF, can not induce large changes in the 
visible brightness of that star. If the asymmetric spectroscopic properties of 
the primary in the orbital plane that are hinted at from the width of the CCF 
are due to (say) mass accretion onto the primary, then the modest 
variations in the light curve indicates that there is not a spot 
on the surface of the primary or a surrounding disk with an effective 
temperature that differs greatly from its surroundings.

\subsection{The Evolutionary Status of PS: Past and Future}

	It is widely accepted that PS experienced 
mass transfer after the primary evolved into contact with its 
critical Roche surface. Evidence to support this is 
the rapid rotation of the secondary due to the accretion of angular 
momentum from the primary, and the CNO surface abundances 
of the stars \citep[e.g.][]{bagandbar1996, linetal2006, linetal2008} 
that are indicative of core processed material that has been brought 
to the surface, as might be expected if the primary had evolved off of the 
main sequence prior to mass transfer. The evidence that the fainter secondary 
is now more massive than the brighter primary at visible wavelengths is 
a further indication that there has been mass transfer. 

	\citet{staetal2024} suggest that PS may be a highly evolved 
system, in which the primary has lost most of its mass 
and is now a heavily stripped, bloated remnant. Evidence for this comes from 
the radial velocity variations deduced from the Stokes V profile 
measurements made by \citet{gruetal2022}. If K$_2 = 30$ km/sec 
as found by \citet{gruetal2022} then the system mass 
ratio is $\frac{202}{30} = 6.7$. The component masses are then 
M$_1$ sini$^3$ = 2.5 M$_\odot$ and M$_2$sini$^3$ = 16.7 M$_\odot$. 
Adopting $i = 71^o$ as was done in Section 6 then the component masses 
would be roughly 3 and 20M$_\odot$; the latter is close to that 
expected for an O8 main sequence star.

	Based on more recent observations 
than those examined by \citet{gruetal2022}, \citet{staetal2024} 
find that the radial velocity amplitude of the secondary may be even smaller 
than that reported by \cite{gruetal2022}. If this is the case then 
the masses of the primary and secondary would be pushed to even lower 
values than those stated in the previous paragraph. Still, the mass function 
obtained by \citet{sti1987} indicates that the lower mass 
limit for the secondary is 12.6sini$^3$m$_\odot$.

	The lack of large-scale structure in the system light curve 
with a cadence that matches the orbital period indicates that the 
components of PS are not at present in contact with their Roche surfaces. 
This is consistent with PS being a post-Algol 
system. That the primary in early-type systems can detach from its 
Roche surface after mass transfer is predicted by models. 
\citet{penetal2022} model a $30 + 27$M$_{\odot}$ system with a 20 day 
initial period. Following mass transfer, the mass 
losing star eventually de-couples from its Roche surface and is able to regain 
thermal equilibrium.

	The future evolution of PS is difficult to predict given the 
inherent uncertainties in our understanding 
of the early evolution of stars in massive binary 
systems \citep[e.g.][]{beletal2022}. Still, the component separation 
may decrease in the immediate future. Mass that is lost from PS 
due to winds will remove angular momentum from the system, thereby 
shortening the period. Moreover, if the secondary is more massive than 
the primary and has the dominant wind \citep[e.g.][]{wigandgie1992} then 
there will likely be a net transfer of mass from the secondary to the primary, 
and this will also decrease the component separation. 

	One of the stars in PS will eventually evolve 
into contact with its critical Roche surface and induce another 
episode of large-scale mass transfer. This will affect the separation 
of the components, although it is not obvious if the next 
lobe-filling star will be the primary or the secondary. While the secondary 
is now more massive than the primary, it is also rotating at a much faster rate,
and so mixing will extend its evolutionary timescale when compared with that 
expected for a non-rotating star. If there is another mass transfer event 
that is spurred by the evolution of the primary then 
this should cause the separation of the components to increase, 
although this depends on the rate of mass loss from the system 
if the mass transfer is not conservative. 

\subsection{Future Work}

	There is considerable scope for future work, and 
a recommended emphasis is on observations that will 
help to better understand the nature of the secondary. There is 
uncertainty in the amplitude of the velocity curve of the secondary, and 
thus the system masses. Spectra that examine PS through a number of 
sequential cycles would be of interest to chart variations in the 
velocity curve of the secondary, and place limits on the timescale 
of any variations in the amplitude of its orbital velocity curve. 
That the radial velocity amplitudes of the secondary obtained from the 
1922 and 1937 campaign spectra appear to differ indicates that the amplitude 
can change on timescales $\leq 15$ years, although other velocity measurements 
for the secondary \citep[e.g.][]{linetal2008} suggest a shorter 
timescale. An investigation of this nature could also explore possible 
correlations with lines that are associated with stellar winds, such as 
\ion{He}{2} $\lambda 4486$ and lines in the UV that show P Cygni 
profiles. These spectra might also provide information that could be 
used to assess if clumping is a possible cause of the velocity variations.

	The behaviour of H$\gamma$ in the spectrum of 
the primary near phase 0.25 could also be examined with spectra that span a 
number of orbits, with an emphasis on determining if variations in line 
strength and velocity (see the Appendix) are seen with modern detectors. 
As the cyclicity of these variations are not known, then 
this could require monitoring over a longer term. 
If variations in the strength of H$\gamma$ are found 
then it would be of interest to search for 
signs of possible abundance variations at this same phase, as abundance 
spots may be one cause of these variations.

\acknowledgements{It is a pleasure to thank the anonymous referee for 
providing a comprehensive report that greatly improved the manuscript.}

\appendix

\section{Radial Velocities}

	These data were originally recorded to measure radial velocities 
as part of a survey of early-type stars. In this section, we 
discuss velocity measurements extracted from the digitized spectra 
using different techniques, with a goal of 
assessing the velocity information that can be extracted, 
especially with regards to the secondary. 
The spectra recorded in the 1920s are of particular interest, 
as they can be compared with the original measurements that were made 
by \cite{pla1922}.

	Many of the 1.2 meter spectra tend to be of poor quality, 
and so the discussion is restricted to 1.8 meter spectra. 
In addition, as the moderate wavelength resolution of the 1.8 meter spectra 
complicates efforts to isolate features from the secondary, then only 
spectra that sample the phase intervals 0.15 -- 0.35 and 0.65 -- 0.85 are 
considered. These are the points in the orbit where the velocity 
difference between the primary and the secondary is 
largest, thereby enhancing the chances of de-coupling the 
spectroscopic signatures of the stars.

	In Figure 13 it is demonstrated that the spectrum of the secondary 
forms a prominent feature in the CCF, and so cross-correlation provides 
a compelling means of extracting velocity information for the secondary. 
Velocities were measured with the IRAF FXCOR task, using the 
wavelength interval between 4100 and 4600\AA\ .
This interval is where the S/N is highest, and is 
also where there is a concentration of prominent absorption lines, of which 
H$\gamma$ and \ion{He}{1} $\lambda 4471$ are the strongest. 
The model template described in Section 6 was used as the reference 
for the cross-correlation, and experimentation found that the velocities 
are not sensitive to the presence of the DIB near 4400\AA\ . Moreover, 
they did change markedly when the wavelength range was expanded. 
The results are presented in Table 4, while the 
velocities are shown as a function of orbital phase in the upper panel 
of Figure A1. The estimated uncertainties in individual velocity measurements 
vary with data quality, but are typically 5 -- 15 km/sec.

\begin{center}
\begin{deluxetable}{lclrr}
\tablecaption{Radial Velocities Measured Between 4100\AA\ to 4600\AA}
\tablehead{Plate \# & HJD & Phase & V$_1$ & V$_2$ \\
 & & & (km/sec) & (km/sec) \\}
\startdata
6953 & 23053.838 & 0.297 & 182.0 & -37.2 \\
7006 & 23067.849 & 0.27 & 188.0 & -68.5 \\
7138 & 23104.753 & 0.834 & -142.7 & 137.3 \\
7193 & 23109.733 & 0.18  & 191.9 & -101.2 \\
7209 & 23110.632 & 0.242 & 202.2 & -54.0 \\
7213 & 23110.736 & 0.249 & 221.6 & -60.8 \\
7217 & 23111.675 & 0.315 & 218.0 & -54.0 \\
7282 & 23124.689 & 0.219 & 220.4 & -54.0 \\
7301 & 23132.635 & 0.771 & -191.9 & 112.0 \\
7302 & 23132.808 & 0.783 & -166.9 & 138.5 \\
7304 & 23133.633 & 0.84  & -129.9 & 80.8 \\
26236 & 28459.062 & 0.758 & -219.4 & 64.0 \\
26367 & 28494.051 & 0.188 & 182.0 & -37.2 \\
26568 & 28552.854 & 0.272 & 235.6 & -74.2 \\
26649 & 28595.736 & 0.251 & 244.1 & -54.0 \\
26726 & 28630.661 & 0.677 & -100.9 & 182.0 \\
28624 & 29178.027 & 0.699 & -210.6 & 102.2 \\
28992 & 29257.906 & 0.247 & 150.6 & -116.7 \\
A22101 & 29322.704 & 0.748 & -151.4 & 64.0 \\
\enddata
\end{deluxetable}
\end{center}

\begin{figure}
\figurenum{A1}
\epsscale{1.0}
\plotone{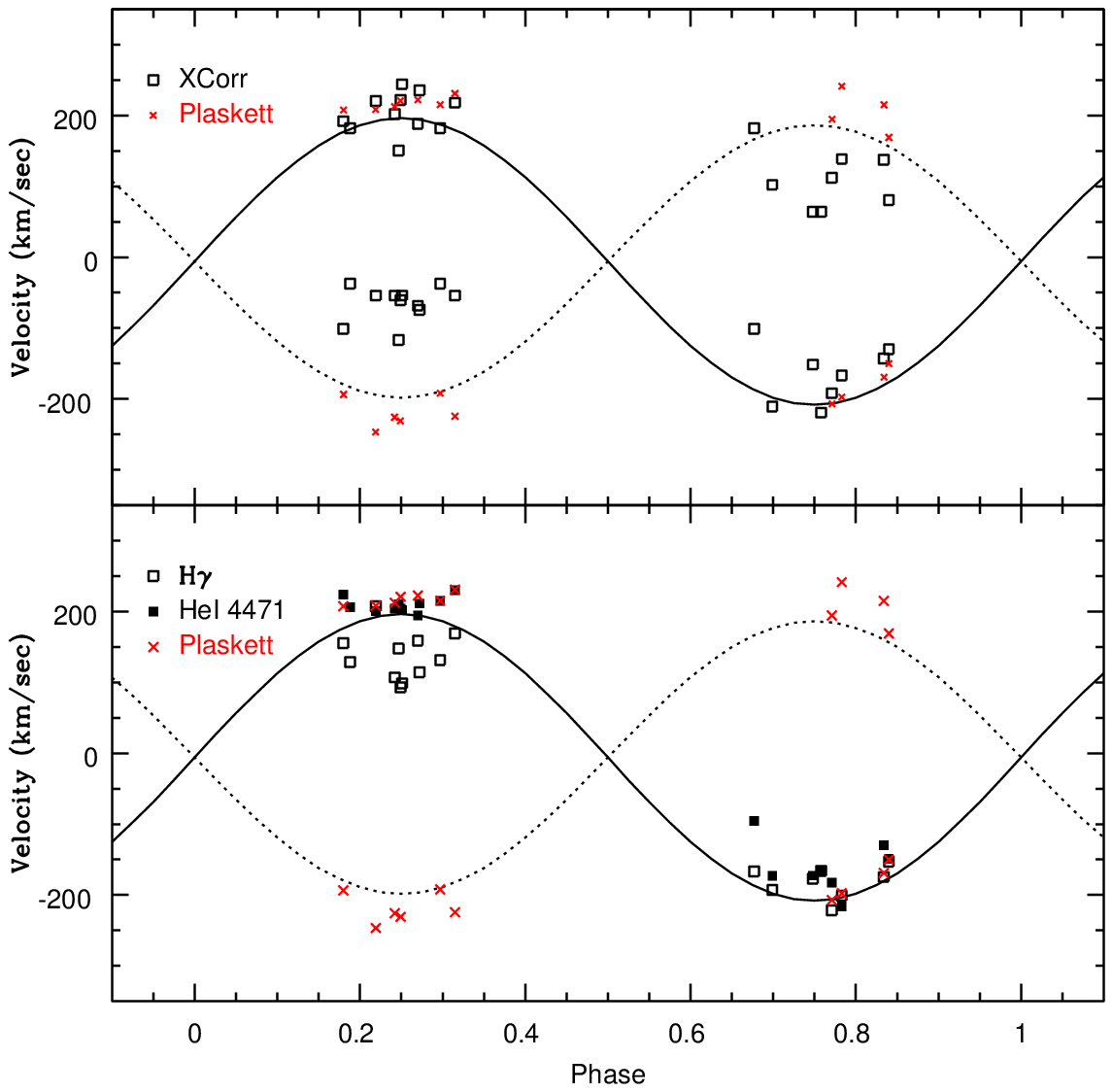}
\caption{Radial velocities measured from digitized 1.8 meter spectra 
using different techniques. The top panel shows CCF-based velocities 
while the lower panel shows velocities of the primary estimated from the cores 
of H$\gamma$ and \ion{He}{1} $\lambda 4471$. The red crosses in both panels 
are the velocities measured by \cite{pla1922} from a number of 
lines and summarized in his 'Table of Velocities and Residuals'. 
The solid and dashed curves are the radial velocity curves 
based on the orbital elements calculated by \citet{linetal2008}.}
\end{figure}

	We estimate the amplitude of the velocity curve 
at each epoch by taking the mean velocity of the primary and 
secondary in each phase interval and then taking the difference. 
The full amplitude of the primary velocity curve is $361 \pm 15$ km/sec 
in the 1920s, and $373 \pm 36$ km/sec in the 1930s. These tend to be 
lower than the full amplitudes found from more recent studies because 
they are averages over a large range of the system orbit. 
The full amplitude of the secondary velocity curve 
from the 1920s measurements is then $179 \pm 16$ km/sec while the 
narrow measurements from the 1930s gives an amplitude $174 \pm 31$ km/sec. The 
uncertainties are $1-\sigma$ formal errors of the mean. 
The ratio of the amplitudes is then $0.48 \pm 0.11$.

	The cores of individual deep lines provide robust velocity 
estimates for the brighter star that are less susceptible to contamination 
from a fainter companion. Velocities were measured from the cores 
of H$\gamma$ and \ion{He}{1} $\lambda 4471$, which are among the most 
prominent features in the 1.8 meter spectra, and the 
results are tabulated in Table 5, where the means of the H$\gamma$ 
and \ion{He}{1} $\lambda 4471$ measurements are in the final column. 
The estimated uncertainties are $\pm 10$ km/sec.
We have not attempted to extract velocities for the secondary from 
line centroids given the heavily blended nature of the lines. The 
H$\gamma$ and \ion{He}{1} $\lambda 4471$ velocities are plotted in the 
lower panel of Figure A1.

\begin{center}
\begin{deluxetable}{lclrrr}
\tablecaption{Radial Velocities Measured From Cores of H$\gamma$ and \ion{He}{1} $\lambda 4471$ }
\tablehead{Plate \# & HJD & Phase & V$_H$ & V$_{He}$ & V$_{Mean}$ \\
 & & & (km/sec) & (km/sec) & (km/sec) \\}
\startdata
6953 & 23053.838 & 0.297 & 131.4 & 214.8  & 173.1 \\
7006 & 23067.849 & 0.27 & 158.9 & 194.27 & 176.6 \\
7138 & 23104.753 & 0.834 & -174.7 & -129.9 & -152.3 \\
7193 & 23109.733 & 0.18  & 155.6 & 224.1 & 189.9 \\
7209 & 23110.632 & 0.242 & 107.0 & 203.8 & 155.4 \\
7213 & 23110.736 & 0.249 & 93.2 & 203.8  & 148.5 \\
7217 & 23111.675 & 0.315 & 168.9 & 230.3 & 199.6 \\
7282 & 23124.689 & 0.219 & 207.7 & 200.8 & 204.3 \\
7301 & 23132.635 & 0.771 & -221.8 & -182.6 & -202.2 \\
7302 & 23132.808 & 0.783 & -201.1 & -216.2 & -208.6 \\
7304 & 23133.633 & 0.84  & -152.8 & -149.2 & -151.0 \\
26236 & 28459.062 & 0.758 & -166.0 & -167.0 & -166.5 \\
26367 & 28494.051 & 0.188 & 128.8 & 206.1 & 167.4 \\
26568 & 28552.854 & 0.272 & 114.3 & 211.2 & 162.7 \\
26649 & 28595.736 & 0.251 & 98.6 & 202.3 & 150.4 \\
26726 & 28630.661 & 0.677 & -166.59 & -95.4 & -131.0 \\
28624 & 29178.027 & 0.699 & -192.6 & -172.7 & -182.7 \\
28992 & 29257.906 & 0.247 & 147.8 & 210.6 & 179.2 \\
A22101 & 29322.704 & 0.748 & -177.1 & -172.6 & -174.8 \\
\enddata
\end{deluxetable}
\end{center}

	The \ion{He}{1} velocities near phase 0.25 have modest scatter 
when compared with their H$\gamma$ counterparts. Moreover, the amplitude 
of the velocity curve defined by \ion{He}{1} $\lambda 4471$ is also 
comparable to that found in previous studies, although there might be a 
slight offset in the system velocity obtained from this line. 
What is most noticeable is that while H$\gamma$ is 
the most prominent feature in these spectra and is 
arguably the best defined, it shows a marked departure 
from the Linder et al. (2008) velocity curve near phase 0.25. At this 
phase the H$\gamma$ velocities are consistently lower than those 
obtained from \ion{He}{1} $\lambda 4471$. This offset holds for 
measurements made from both the 1920s and 1930s spectra. The 
strength of H$\gamma$ at this phase also differs from that at 
other phases (e.g. Section 4.2). The agreement between 
the H$\gamma$ and \ion{He}{1} velocities is better near phase 0.75. 
We note that \cite{pla1922} stated that H$\gamma$ tended to give 'good 
velocity measures', although even some of these were 'somewhat discrepant'. 

	In summary, the comparisons made in this section indicate 
that velocities made from the core of \ion{He}{1} $\lambda 4471$ 
in the digitized spectra match those originally 
obtained by \citet{pla1922}, and track velocity variations 
that are based on the orbital elements found by \cite{linetal2008}. 
The velocity measurements for the primary star made 
directly from the plates by \cite{pla1922} 
thus appear to be consistent in quality with those obtained 
from \ion{He}{1} $\lambda 4471$ in the digitized spectra. As for measurements 
made from CCFs, we find that the full amplitude of the velocity 
curve of the secondary is substantially smaller than found by \cite{pla1922}. 
This argues for a mass ratio that is substantially different from near unity.

\end{document}